\documentclass[prb,twocolumn,showpacs,floatfix]{revtex4-1}
\usepackage{graphicx}
\usepackage{color}
\usepackage{amssymb}
\usepackage{amsmath}
\usepackage{amsfonts}
\usepackage{bbm}
\usepackage{xspace}
\usepackage{url}
\usepackage{subfigure}
\usepackage{float}

\newcommand{\ii}{\mathrm{i}}
\newcommand{\ee}{\mathrm{e}}

\begin{document} 
\title{Topological insulators in random  potentials}
%
\author{Andreas Pieper and Holger Fehske}
\affiliation{Institut f\"ur Physik, Ernst-Moritz-Arndt-Universit\"at
  Greifswald, 17487 Greifswald, Germany}
\date{\today}
\begin{abstract}
We investigate the effects of magnetic and nonmagnetic impurities  on the two-dimensional surface states of three-dimensional topological insulators (TIs).  Modeling weak and strong TIs using a generic four-band Hamiltonian, which allows for a breaking of inversion and time-reversal symmetries and takes into account random local potentials as well as the Zeeman and orbital effects of external magnetic fields, we compute the local density of states, the single-particle spectral function, and the conductance for a (contacted) slab geometry by  numerically exact techniques based on kernel polynomial expansion and Green's function approaches. We show that bulk disorder refills the suface-state Dirac gap induced by a homogeneous magnetic field with states, whereas orbital (Peierls-phase) disorder perserves the gap feature.  The former effect is more pronounced in weak TIs than in strong TIs. At moderate randomness, disorder-induced conducting channels appear in the surface layer, promoting diffusive metallicity.  Random Zeeman fields rapidly destroy any   conducting surface states. Imprinting quantum dots on a TI's surface, we demonstrate that  carrier transport  can be easily tuned by varying the gate voltage, even to the point where quasi-bound dot states may appear.
\end{abstract}

\maketitle
\section{Introduction}
Topological insulators (TIs) are novel states of quantum matter which are equally important for both fundamental solid-state physics research and technological applications~\cite{HK10,KWBRBMQZ07,CBPF11}. The remarkable properties of three-dimenensional (3D)  TIs result from the particular topology of their band structure exhibiting gapped bulk and gapless linearly dispersed  Dirac surface states~\cite{Xiea09,Chea09}.  Most notably, the metallic (spin-polarized) surface states  are largely robust against the influence of non-magnetic disorder. Depending on the degree of this robustness  a distinction is drawn between weak and strong 3D ${\mathbb Z}_2$ TIs, where bulk-surface correspondence implies that weak TIs feature none or an even number of helical Dirac cones while strong TIs manifest a single Dirac cone~\cite{FKM07,FK07,RKS12}.  In the weakly disordered regime, the surface states of weak TIs have an internal structure and are either robust or ``defeated'' by disorder~\cite{KOI13}. The conducting surface states of strong TIs are topologically protected against localization due to their helical nature; here the spin-momentum locking suppresses backward scattering as in graphene~\cite{ANS98,KOI13,BR10}. Surface disorder with a strength comparable to or larger than the bulk band gap will destroy the Dirac cone, leading first to diffusive metallic behavior and then to Anderson localization at the surface~\cite{SFFV12,KOIH14}.  

\begin{figure}[b]
\includegraphics[width=0.8\linewidth]{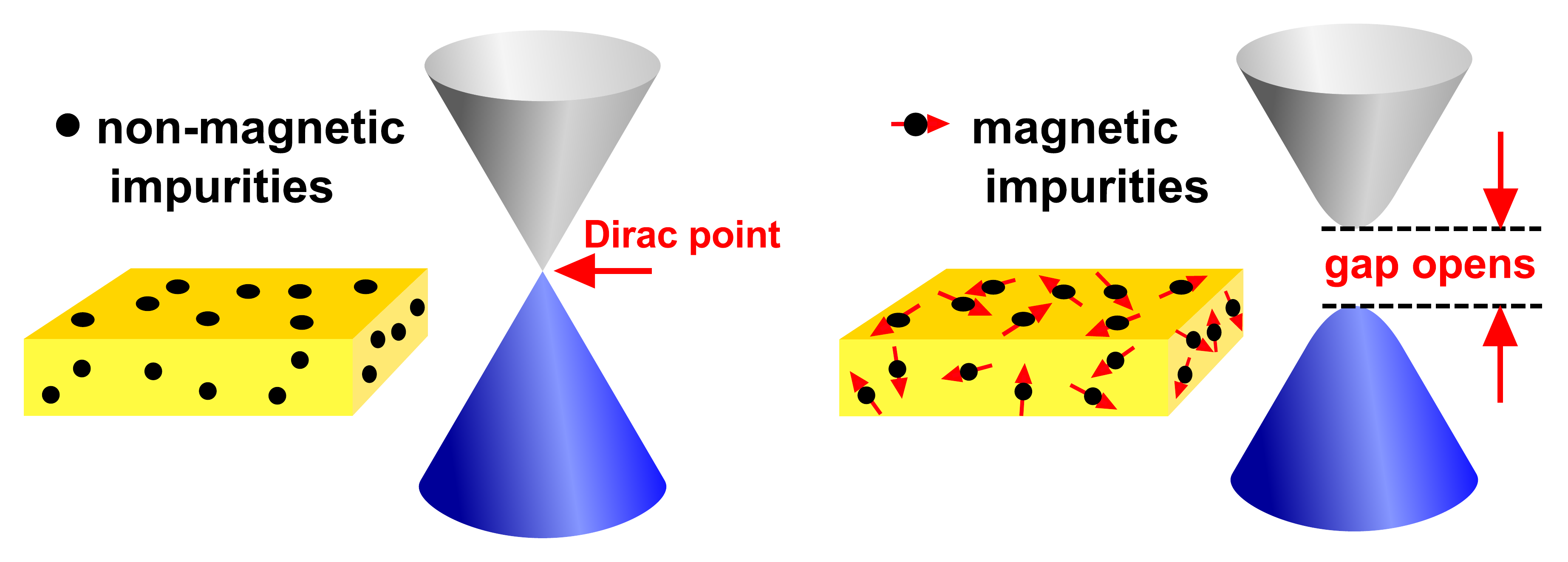}
\caption{
(Color online)
Schematic picture of a topological insulator with non-magnetic (left) and magnetic (right) impurities.
The non-magnetically doped TI has a Dirac point connecting the occupied lower and unoccupied surface states as in the undoped case. In magnetically doped TI a gap separates the upper and lower Dirac cones~\cite{Chea10}.
}
\label{fig1}
\end{figure}

Another way to affect the robust surface metallicity of 3D TIs is to break time-reversal symmetry. This can be achieved by placing magnetic dopants into 3D TI.  As a result insulating massive Dirac electron states will be formed, with striking topological features (see Fig.~\ref{fig1}). Such states have been observed in recent angle-resolved photoemission experiments on magnetically doped dibismuth triselenide with a Dirac gap and a Fermi energy tuned into this surface-state gap~\cite{Chea10}. The effects of magnetic impurities  solely on the surface of a TI have been studied with the intention of a gap opening, which certainly is a promising pathway for functionalizing a TI~\cite{LLXQZ09,BBF15}. 
The idea is that the Dirac electronic states mediate a Ruderman-Kittel-Kasuya-Yosida interaction among the impurities which is ferromagnetic and therefore will produce a net (average Zeeman) magnetic field that gaps the TI's surface states. Moreover, with a view to spintronics applications, specific time-reversal-breaking potentials have been explored that constitute magnetically defined qubits by confining topological surface states into quantum wires or quantum dots~\cite{FL13}. Besides, we like to point out that  
applying magnetic fields  will induce one-dimensional  edge channels which strongly influence the quantum Hall physics of 3D TIs~\cite{SRAF12}.   

Motivated by this situation, in this work, we examine the rather complex interplay of topology and disorder in 3D TIs~\cite{SM09,LLXQZ09,BB10,SFFV12,LRS12,RKS12,MBM12,KOI13,KOIH14,SB14,Liea15,SOK15,BBF15}. To understand the effect of both nonmagnetic and magnetic impurities, we study a minimal four-band Hamiltonian that allows us to introduce and control a gap into the Dirac cone by a homogeneous magnetic field and, in addition,  that includes random on-site potentials as well as  the orbital and Zeeman effects of spatially fluctuating magnetic  fields.  Using exact numerical techniques, we analyze  ground-state, spectral and transport properties of this Hamiltonian, which may realize, on a (contacted) slab geometry, distinct topological phases in different parameter regimes.  We furthermore will show that we can selectively induce states in  the surface Dirac gap, just by imprinting on the TI's surface a gate-defined quantum dot, and thereby can manipulate the surface
currents. In the course of our investigations we will focus on the strong TI case;  selected results, however, will be contrasted 
by the corresponding ones for a weak TI.

\section{Model and method} 
\begin{figure}[th]
\centering
\includegraphics[width=0.8\linewidth]{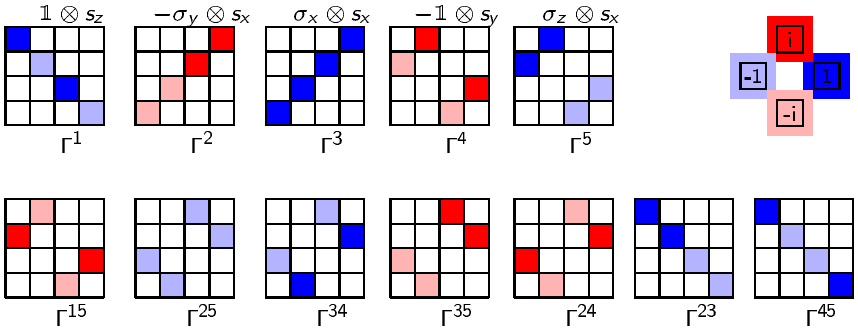} 
\caption{
(Color online)
Schematic representation of the on-site (orbital) matrix elements of the Hamiltonian~\eqref{eq:TB_H} with \eqref{eq:TB_H_BZ}, according to the five Dirac matrices $\Gamma^a$, and their ten commutators $\Gamma^{ab}=[\Gamma^a,\Gamma^b]/2i$, which satisfy the Clifford  algebra, $\{\Gamma^a,\Gamma^b\}= 2\delta_{a,b} \Gamma^0$ ,with $\Gamma^0$ being the identity. Here, $\sigma_i$ and $s_i$ denote the Pauli matrices. 
}
\label{fig2}
\end{figure}

\begin{figure}[th]
\centering
\includegraphics[width=0.8\linewidth]{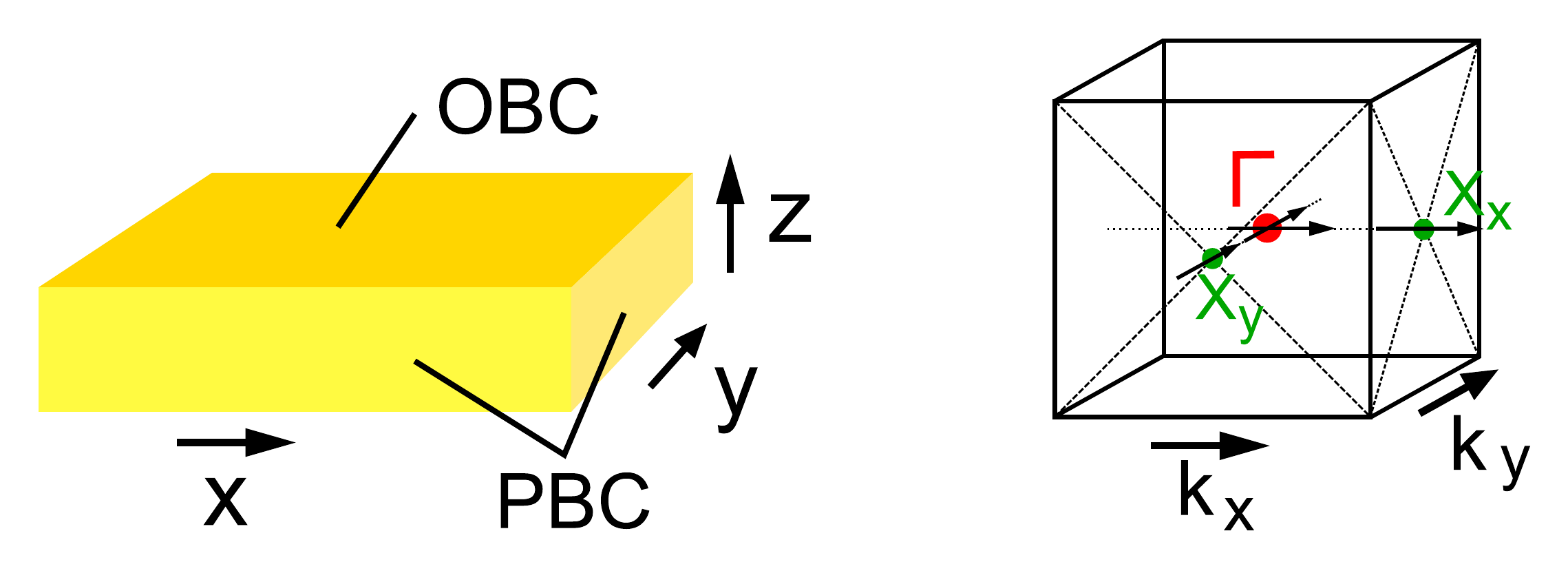}
\includegraphics[width=0.98\linewidth]{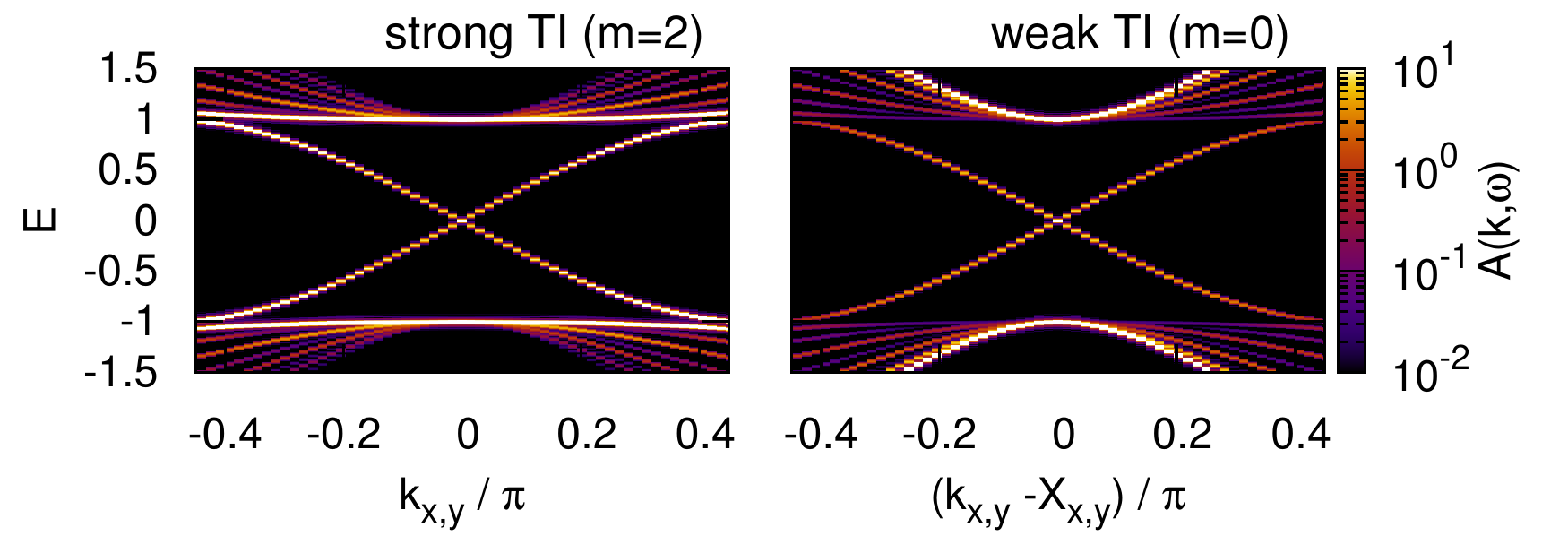} 
\includegraphics[width=0.98\linewidth]{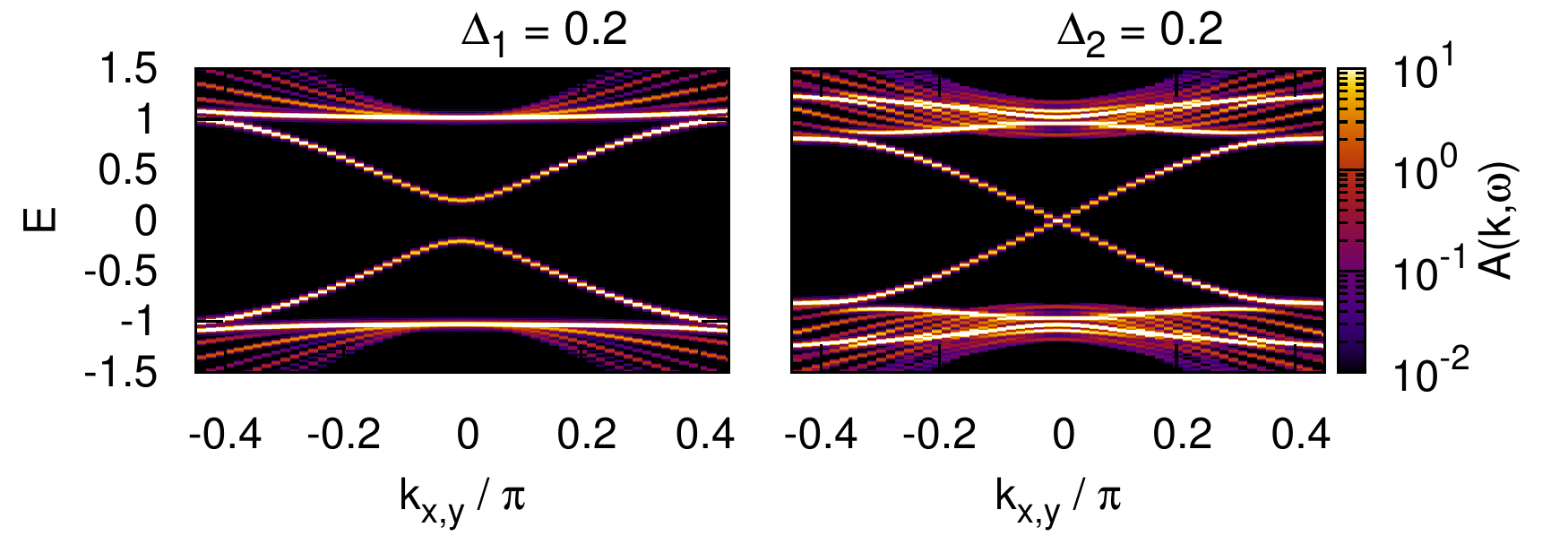} 
\includegraphics[width=0.98\linewidth]{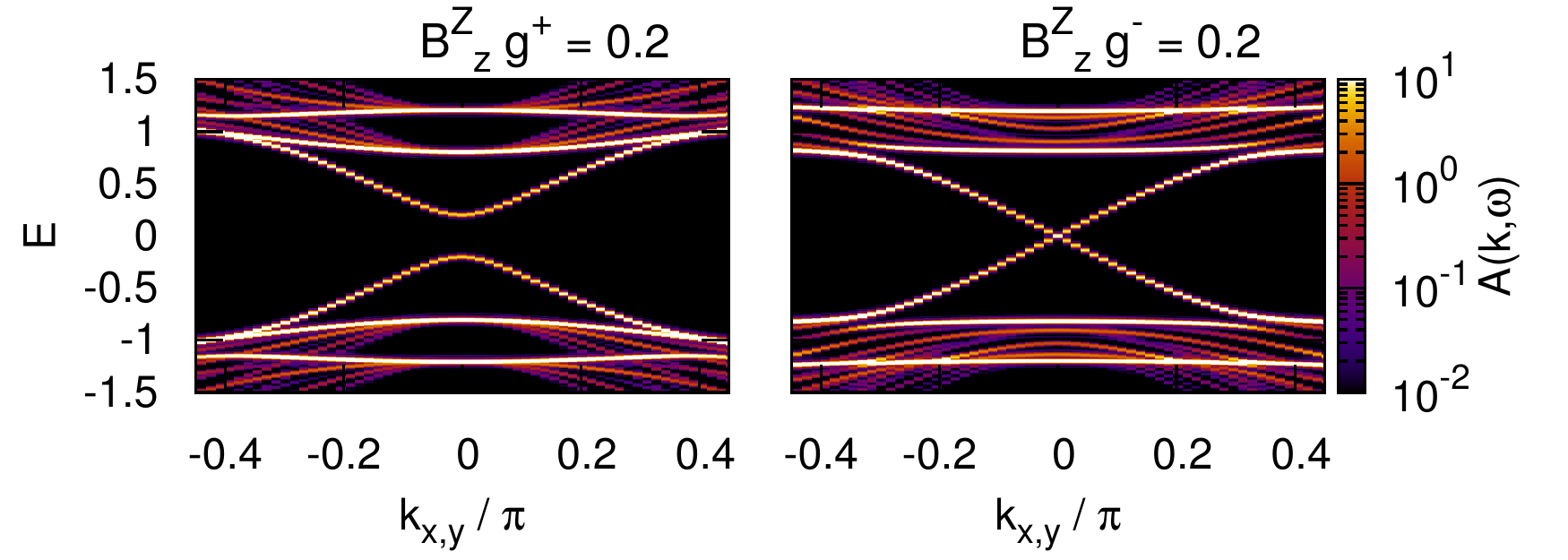} 
\caption{
(Color online) Top: Schematic representation of the TI slab geometry with periodic boundary conditions (PBC) in the $x$ and $y$ directions and open boundary conditions (OBC) in the $z$-direction  [left],  as well as of the bulk Brillouin zone [right]. Bottom: Band structure of a regular TI for all sorts of different cases. Here, the top row shows strong and weak
TIs with one Dirac cone per surface and two Dirac cones per site, respectively, where, $\Delta_{1,2}=0$, $B^Z_{n,j}=0$, and $\tau_{n,j}=0$. Assuming that the magnetic field (moments of the magnetic impurities) is aligned along the $z$-axis, the middle and bottom rows show that for $\Delta_1 \neq 0$ or $g^+ B^Z_z \neq 0$ a gap opens at the at the $\Gamma$-point [left]. $\Delta_2$ and $g^- B^Z_z$, on the other hand, leave the Dirac cone(s) unaffected [right].
}
\label{fig3}
\end{figure}

The regular version of the theoretical model considered below was introduced for TIs with cubic lattice structure, inspired by the orbitals of strained 3D HgTe or the insulators of the $\rm Bi_2Se_3$ family~\cite{FKM07,DHQFZ08,LQZDFZ10,SRAF12}. The corresponding four-band  Hamiltonian is conveniently expressed using the identity $\Gamma^0$, the Dirac matrices $\Gamma^a$, and the commutators $\Gamma^{ab}$ (see Fig.~\ref{fig2} for their matrix elements):
\begin{eqnarray}\label{eq:TB_H}
         H = -t\sum_{n, j} \Big(  \Psi_{n+\hat \ee_j}^\dagger
                  \frac{\Gamma^1 - \ii \Gamma^{j+1} }{2}\,
                 \ee^{\ii \tau_{n,j}} \,
                \Psi_{n} + \text{H.c.}\Big) &&\\
              + \sum_n\Psi_{n}^\dagger \left(
                      m  \,     \Gamma^1
                    + \Delta_1 \Gamma^5
                    + \Delta_2 \Gamma^{15}
                    + V_n \mathbbm{1}
                   \right) {\Psi_{n}}
              + H^{Z}\nonumber 
\end{eqnarray}
with
\begin{eqnarray}\label{eq:TB_H_BZ}
         &H^{Z} =  \sum_{n} \Psi_{n}^\dagger \left[ 
                      - B^{Z}_{n,x} \left(  g^+ \Gamma^{25} + g^- \Gamma^{34} \right)\right.\\
                      & \hspace*{2.2cm}- B^{Z}_{n,y}  \left(  g^+ \Gamma^{35}
- g^- \Gamma^{24} \right)\nonumber\\&\hspace*{2.9cm} \left.
                      + B^{Z}_{n,z}  \left(  g^+ \Gamma^{23} + g^- \Gamma^{45} \right)
                   \right] {\Psi_{n}} \,.\nonumber
\end{eqnarray}
$\Psi_n$ is a four-component spinor at site $n$. In \eqref{eq:TB_H}, nearest-neighbor particle transfer takes place with amplitude $t$, where the orbital effects of an external magnetic field are considered by the Peierls factor $\ee^{\ii \tau_{n,j}}$, acting on the link $n \to n+\vec{e}_{j}$ $(j=1,2,3)$. The parameter $m$ can be used to tune the band structure: For $|m|<t,$ a weak TI with two Dirac cones per surface arises, whereas for $t<|m|<3t$, a strong TI  results, with a single Dirac cone per surface (see Fig.~\ref{fig3}). In the case that $|m|>3t$ we have a conventional band insulator. External magnetic fields
cause finite $\Delta_1$ and $\Delta_2$, which will break the inversion symmetry. $\Delta_1$, in addition, breaks the time-inversion symmetry. The Zeeman effect of the magnetic field  is described by $H^{Z}$, where $\vec{B}^{Z}=\mu_{\rm B}\vec{B}$, and $g^\pm$ are linear combinations of the $g$ factors of the $E1$ and $LH$ subbands~\cite{SRAF12}. Most notably, both $\Delta_1$ and $B^{Z}_z g^+$--but not $\Delta_2$, $B^{Z}_z g^-$--open a gap in the band structure (see Fig.~\ref{fig3}; there, and in what follows, we have  set $t=1$, fixing the energy scale).

We implement the effect of nonmagnetic impurities by random on-site potentials  $V_n$.
For (quenched) bulk Anderson  disorder,  $V_n$ are drawn from a uniform probability distribution, i.e., $p[V_n]=\tfrac{1}{\gamma}\theta(\tfrac{\gamma}{2}-|V_n|)$. Surface disorder is realized if $n$ belongs to the lateral faces of the sample ($V=0$ otherwise). Note that we can use $V_n$ as well in order to establish gate-defined
quantum structures, such as electrostatically defined quantum dots  or quantum dot superlattices~\cite{PHWF14}. For example, $V_n=V_{dot}\Theta(R-|\vec{r}_n-\vec{r}_{dot}|)$ imprints, for a suitable choice of $\vec{r}_n$ and $\vec{r}_{dot}$, a circular region on the surface, whereby the additional potential $V_{dot}$ will scatter or possibly even trap the charge carriers. 

The effect of (surface) magnetic impurities has been previously studied, in the framework of 
a spin exchange Hamiltonian~\cite{LLXQZ09} or 3D tight-binding and effective continuum surface models~\cite{BBF15}, for the cases  of an isolated impurity, uniformly distributed impurities and quenched magnetic disorder. In those cases, mean-field, renormalization group~\cite{LLXQZ09}, and  $T$-matrix approaches~\cite{BBF15} 
were used. Here we model magnetic impurities on the surface just as as in the bulk  by random Peierls phases $p[\tau_{n,j}]=\tfrac{1}{\nu}\theta(\tfrac{\nu}{2}-|\tau_{n,j}|)$, or random magnetic fields, with $|\vec{B}^{Z}_n - b \vec{e}_{z}|$ uniformly distributed in the interval $[0,\beta]$, where $\langle  \vec{B}^{Z}\rangle=(0,0,b)$.
  
In order to characterize the ground-state and spectral properties of the model Hamiltonian, we analyze in what follows the density of states (DOS), the local density of states (LDOS), and the single-particle spectral function $A(\vec k, \omega)$ for a given sample geometry and disorder realization. These quantities are given as  
\begin{equation}
  {\rm DOS}(\omega) = \sum_{m=1}^{4N} \delta(\omega - \omega_m)\,,
\label{DOS}
\end{equation}
\begin{equation}
  {\rm LDOS}(\vec {r}_n,\omega) = \sum_{s=1}^4 \sum_{m=1}^{4N}  | \langle m | \Psi(\vec {r}_n,s) \rangle |^2 \delta(\omega - \omega_m)\,,
\label{LDOS}
\end{equation} 
and
\begin{equation}
  A(\vec k, \omega) = \sum_{s=1}^4 \sum_{m=1}^{4N} | \langle m | \Psi(\vec k,s) \rangle |^2 \delta(\omega - \omega_m)\,,
\end{equation}
where $\omega$ is the energy (frequency),  $\vec{r}_n$ denotes the position vector of Wannier site $n$, $\vec{k}$ is the wave vector (crystal momentum) in Fourier space, $N$ labels the number of lattice sites, and $|m\rangle$ designates  the single-particle eigenstates  with energies $\omega_m$. The four-component (ket-) 
spinor $|\Psi(\vec k,s) \rangle$ ($s=1\ldots4$) can be used to construct a Bloch state, just by performing the scalar product  with the canonical (bra-) basis vectors of position and band index spaces~\cite{SFFV12}.  The conductance of a lead-contacted TI, can be obtained, in the limit of a vanishing bias voltage within the standard Landauer--B\"uttiker approach, as  
\begin{equation}\label{eq:cond}
G=\frac{e^2}{h}\sum_{ l\in L, r \in R} | S_{l,r} |^2\,,
\end{equation}
where  $S_{l,r}$ is the scattering matrix between all open (i.e., active) $L$--$R$ lead channels~\cite{Dat95}.

For the calculation of the (L)DOS and spectral function we employ state-of-the-art exact diagonalization techniques~\cite{WF08a},  combined with a Chebyshev expansion and kernel polynomial methods~\cite{WF08,WWAF06}. To evaluate numerically the conductance for a two-terminal setup,  we use the `KWANT' software package~\cite{GWAW14}.

\section{Numerical results and discussion}
\subsection{TI with  nonmagnetic disorder}
\begin{figure}[t]
\includegraphics[width=1.0\linewidth]{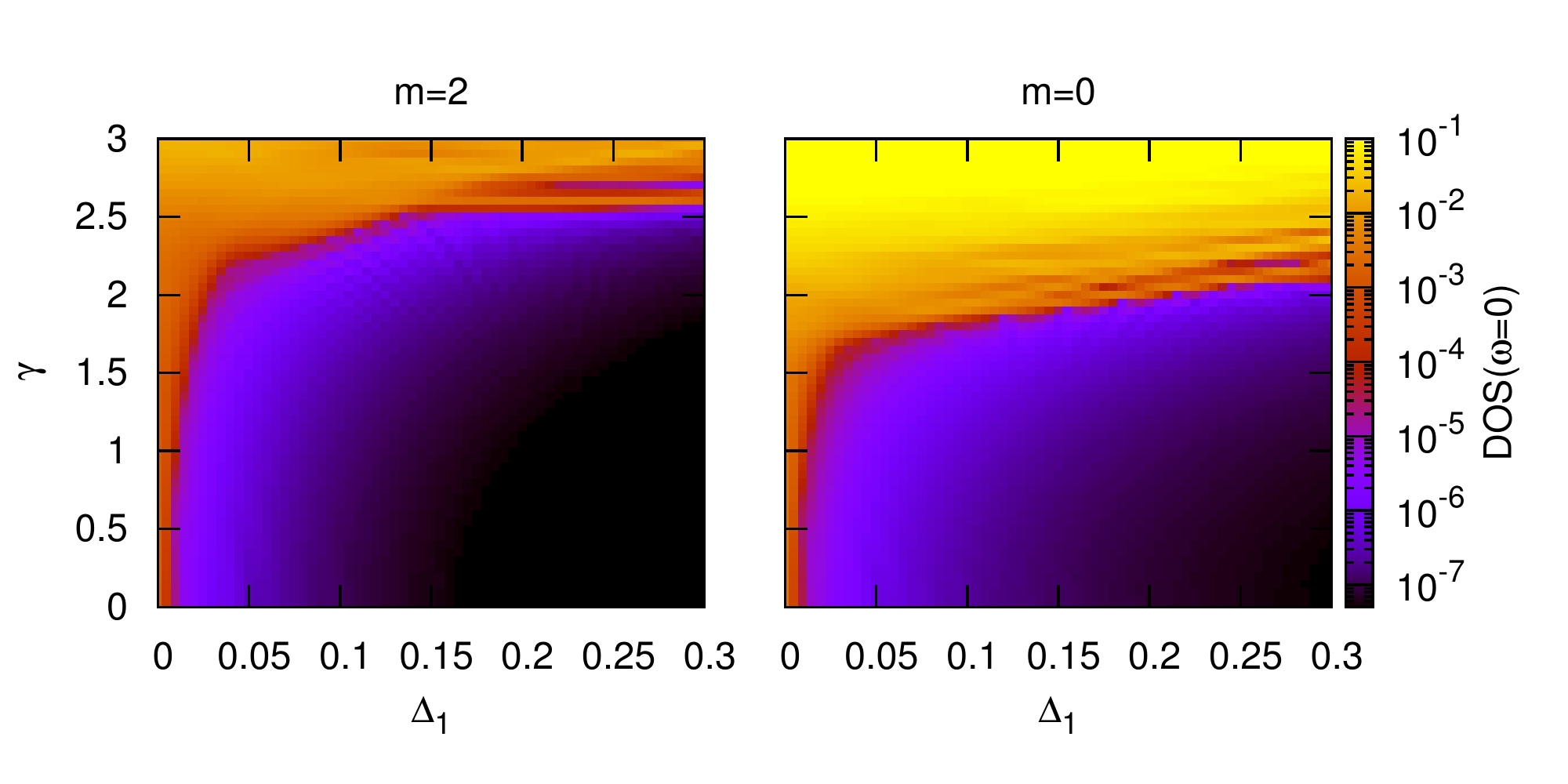} 
\caption{
(Color online) Magnitude of the DOS at the band center, $\omega=0$, as a function of  the gap parameter $\Delta_1$ and 
the strength of the bulk disorder $\gamma$, for a strong [left] and weak [right] TIs with $512\times 512 \times 10$ sites.
 OBC (PBC) are applied in the $z$ direction ($x$ and $y$ directions). We note that the finite DOS at $\Delta_1=0$ (very small  $\Delta_1$) is 
 due to finite-size effects, including the finite number (2048) of Chebyshev moments used in the KPM calculation and the finite KPM resolution 
 with variance of $\sigma^2=0.01^2$.}
\label{fig4}
\end{figure}

We begin by investigating the effect of nonmagnetic impurities in the bulk on the electronic properties of weak and strong TIs, in the case 
where the midgap Dirac cone formed by the surface states is split by a finite $\Delta_1$, caused, e.g., by an external magnetic field (cf. Fig.~\ref{fig3}). In order to simulate an infinite (bounded) system we apply periodic boundary conditions (PBC) in the $x$ and $y$ directions, and open baoudary conditions (OBC) in the  $z$ direction. As a start,  in Eq.~\eqref{eq:TB_H} the Zeeman term is neglected and we consider random on-site potentials in the bulk with $p[V_n]$.  Figure~\ref{fig4} gives the DOS at $\omega=0$, i.e., in the band center of the spectrum (band gap), depending  on both the magnitude of $\Delta_1$ and the strength of disorder  $\gamma$. Although the DOS  displayed corresponds to a  single sample (disorder realization), the contour plot is nevertheless characteristic of the system's behavior because of the large system size and the PBC being used. This has been checked by calculating the DOS data for a couple of samples.  The plots show that as  the disorder strengthincreases , more and more electronic states pop up at the band center. This effect is more strong pronounced, i.e., the gap fills more readily, for weak than for strong TIs. Of course,  the DOS at the band center saturates if $\gamma$ reaches the magnitude of the bulk band gap.

\begin{figure}[t]
\includegraphics[width=0.75\linewidth]{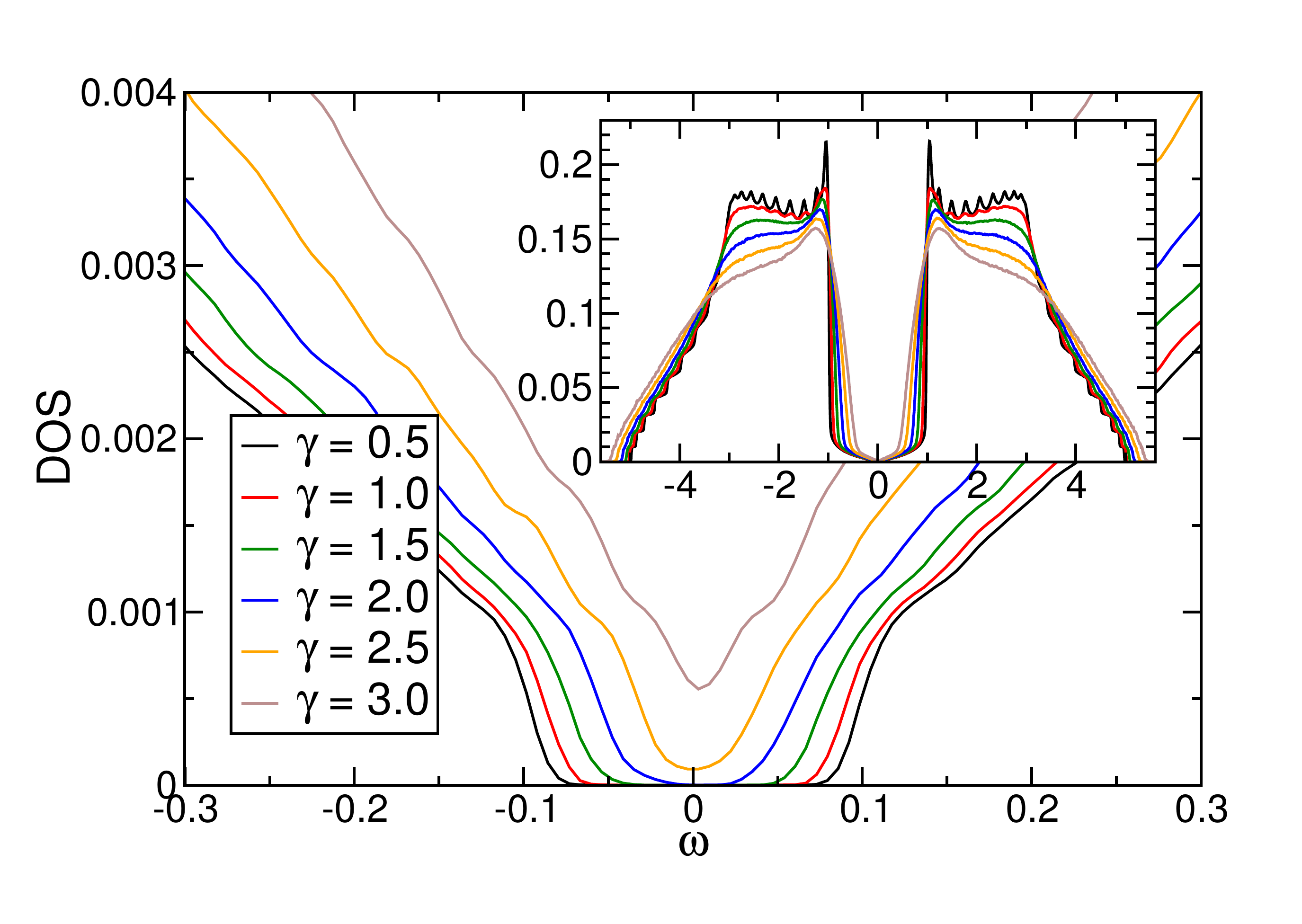}
\includegraphics[width=0.89\linewidth]{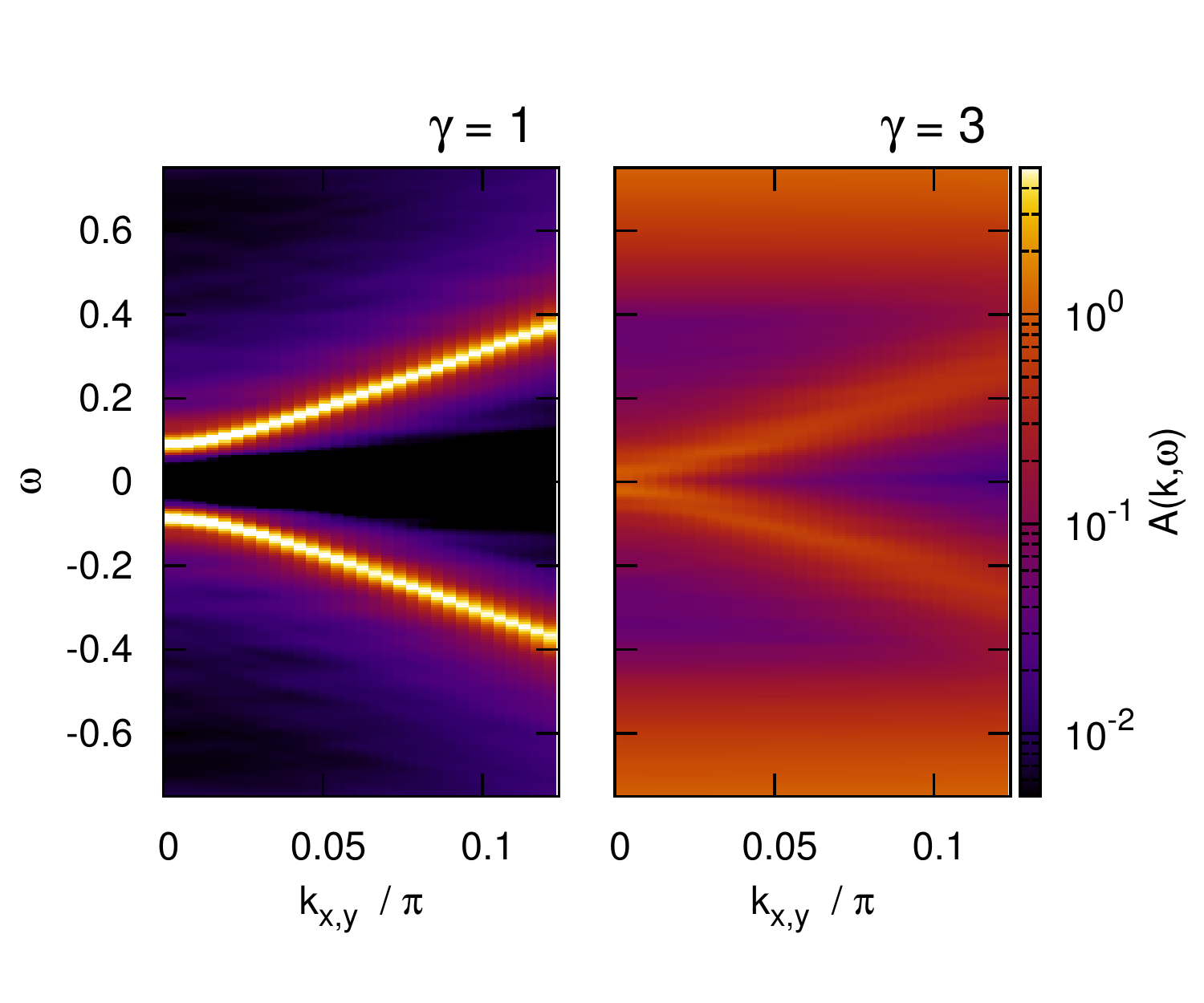} 
\caption{
(Color online)
 Top: Density of states  for bulk disordered TIs. 
Bottom:  Corresponding momentum- and energy-resolved single-particle spectral function.
 The finite $\Delta_1 = 0.1$ leads to a gapped bare band structure. 
 Since a finite $\Delta_2$ does not change the results qualitatively, we show results for only $\Delta_2=0$. 
 The Zeeman term~\eqref{eq:TB_H_BZ} is neglected. 
}
\label{fig5}
\end{figure}

\begin{figure}[t]
\centering
\includegraphics[width=0.98\linewidth]{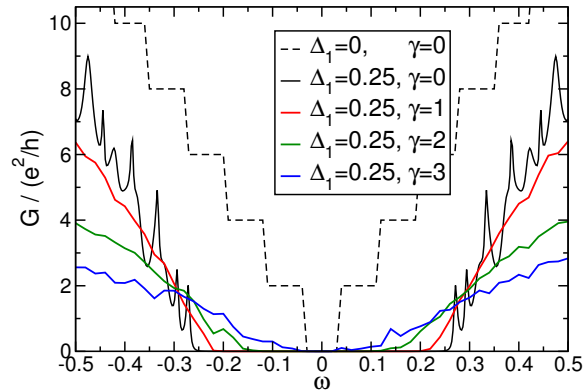} 
\caption{
(Color online) 
Two-terminal conductance $G(\omega)$ of a bulk disordered strong TI ($m=2$) slab with $32\times 32 \times 6$ sites and OBC.  
For the leads we assume $\Delta_1 = 0$; the TI is contacted at $x=0$ and $x=33$. The results shown are averages over ten disorder realizations.}
\label{fig6}
\end{figure}

The entire DOS is shown in the top  panel of Fig.~\ref{fig5} for the strong TI case ($m=2$). We see how bulk disorder induces electron states in the
band gap  region; even so, a pseudo-gap feature remains for weak to intermediate disorder strengths. At the same time any finite-size effects in the bulk-state DOS are washed out by disorder.  The momentum-resolved single-particle spectral function depicted in the bottom panels provides detailed information about the way  disorder influences bulk and surface states.  Measured by angle-resolved photoemission spectroscopy, this
quantity reflects the electronic band structure of disordered TIs. Interestingly, the surface states are almost unaffected by noticeable disorder $\gamma=1$, even though the Dirac cone  is destroyed by $\Delta_1$. For weak disorder the spectral weight transfer into the gap is small. Clearly, the gap closes if the on-site disorder exceeds a certain critical value, but note that $\gamma$ has to be much larger than $\Delta_1$ in order to close the gap. This indicates that the main gap-closing mechanism is the transfer of bulk states into the gap, not the blurring of surface states.

\begin{figure}[h]
\centering
\includegraphics[width=0.52\linewidth]{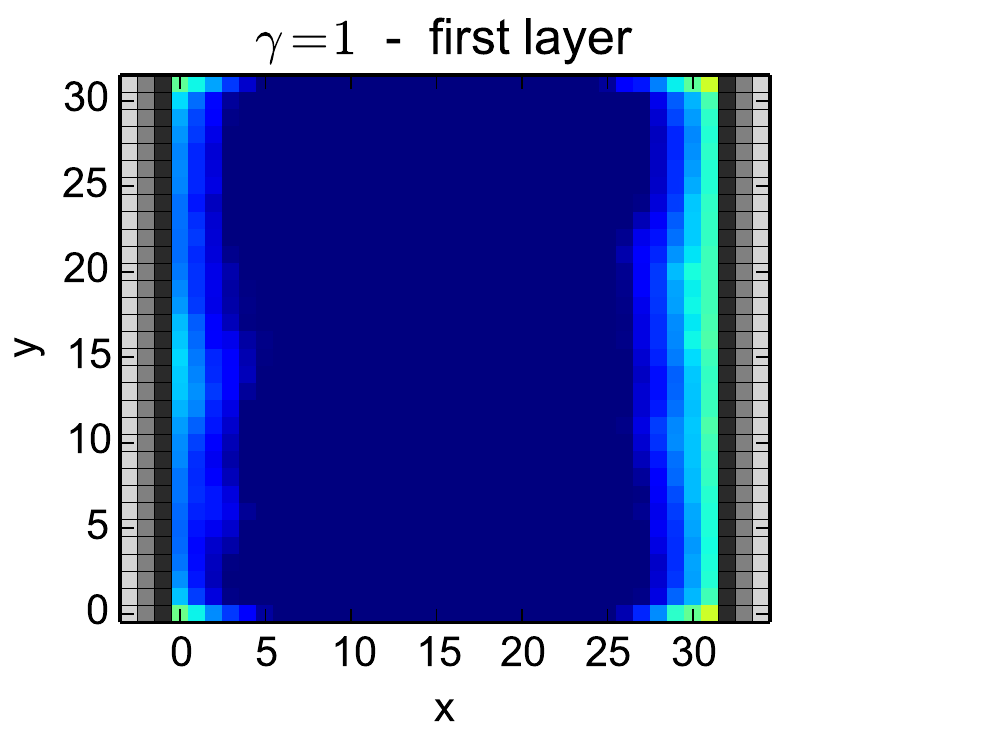} \hspace{-1cm}
\includegraphics[width=0.52\linewidth]{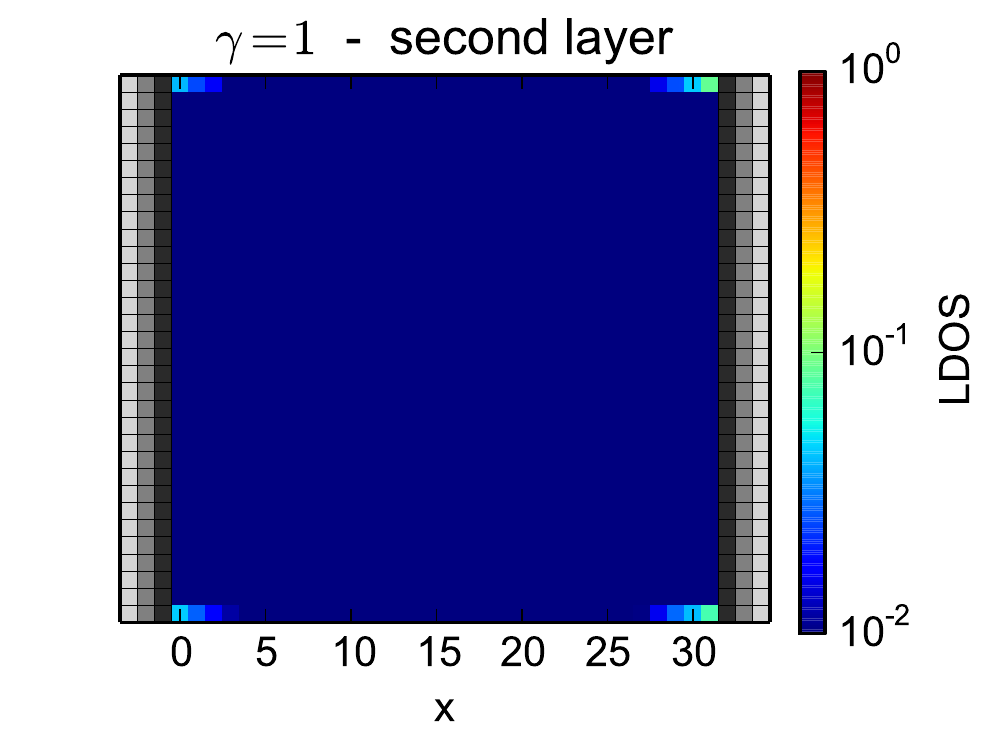}  \\
\includegraphics[width=0.52\linewidth]{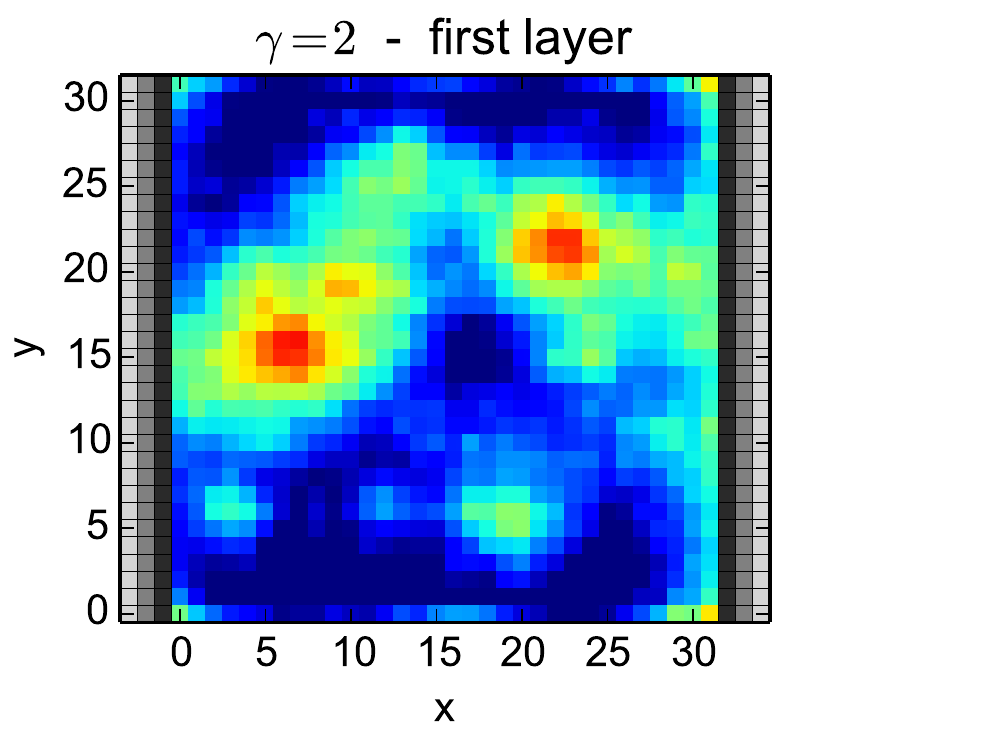} \hspace{-1cm}
\includegraphics[width=0.52\linewidth]{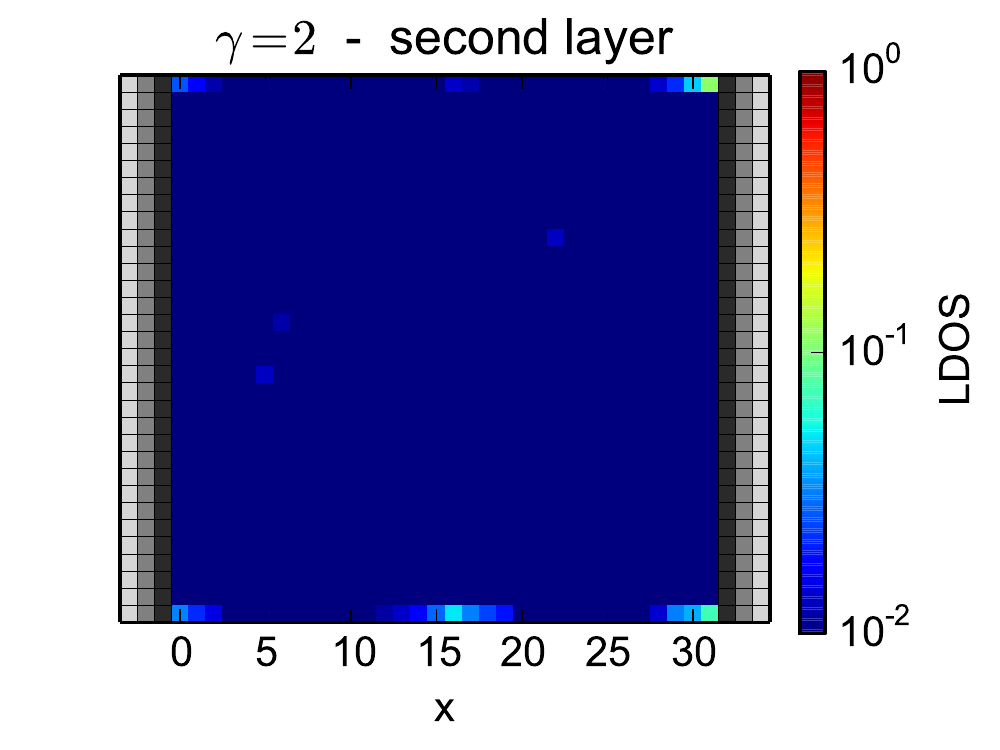}  \\
\includegraphics[width=0.52\linewidth]{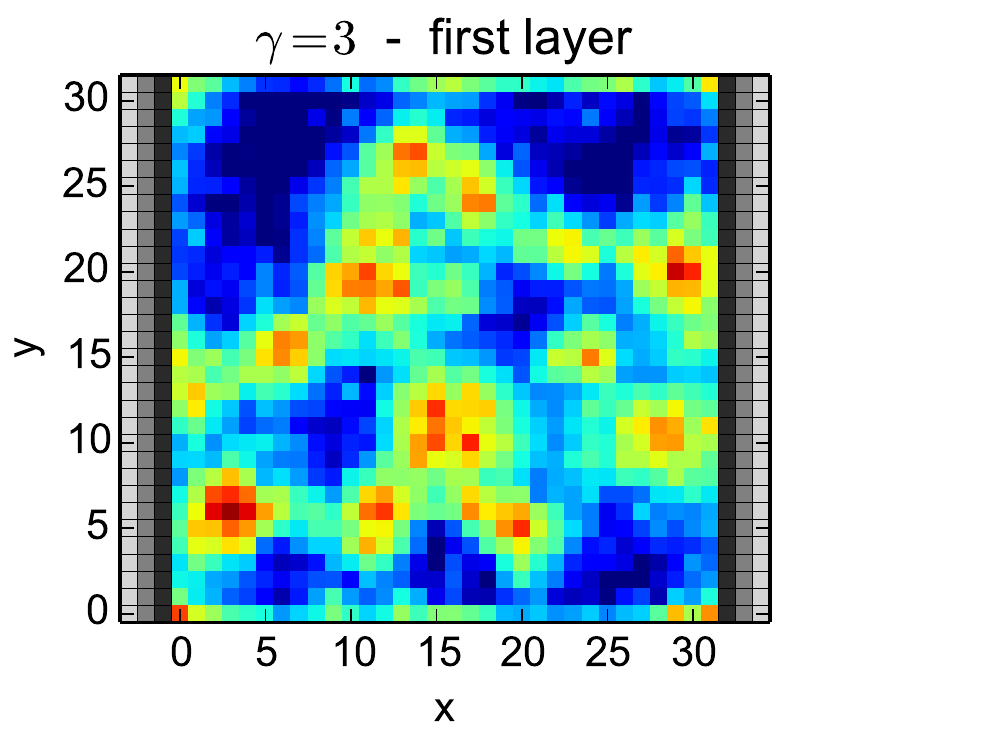} \hspace{-1cm}
\includegraphics[width=0.52\linewidth]{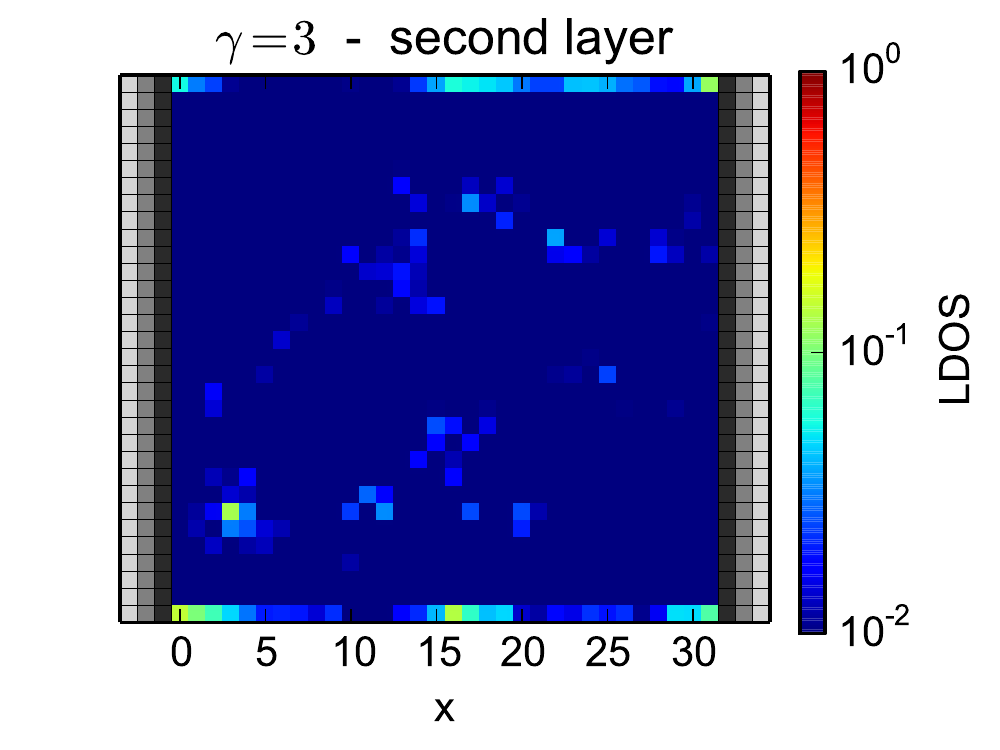}  \\
\caption{
(Color online) Spatially resolved LDOS at $\omega = 0.2$ on the surface [left] and within the first inward layer [right]   
of a single disordered strong TI ($m=2$) sample with $\gamma=1$, 2, and 3 [from top to bottom]. Here $\Delta_1 = 0.25$.  
}
\label{fig7}
\end{figure}

In order to discuss the transport properties of strong TIs we now consider a slab geometry with contacts. 
The leads are assumed to have a gapless band structure, while in the TI sample a gap is induced by $\Delta_1$.   
As can be seen from Fig.~\ref{fig6}, this  inherently changes the conductance of the system without disorder: While one has a finite size gap around $\omega=0$ and a stepwise   enhancement of the conductance when more and more conducting channels contribute to $G$ increasing $\omega$ for  $\Delta_1=0$ (homogeneous system), one finds, in addition the gap due to $\Delta_1$, a spiky $G$ which can be attributed to resonances of the
finite system sandwiched between half-infinite leads~\cite{PSWF13} (see Fig.~\ref{fig7}).  Disorder suppresses these conductance fluctuations of the 
clean lead-TI-lead junction and thereby reduces the conductance within the bulk-state region.  On the other hand,  inside the ($\gamma=0$) gap region, new disorder-induced surface states appear which give rise to diffusive transport, thereby enhancing  the conductance. 

Further information about the nature of these states is obtained from the LDOS displayed in Fig.~\ref{fig7}.  Data are given at $\omega=0.2$, i.e., near the edge of the $\Delta_1$--gap, for the first (surface) and second (first inward) layers of the contacted strong TI sample. We observe, first of all, a finite LDOS in the TI sample near the contacts. This is a pure boundary effect which significantly affects  only the surface layer. More remarkably, bulk disorder induces states which are preferentially localized at the surface and are "self-organized" in such a way  that conducting paths evolve on the surface between the leads. In the first instance this tendency continues if the disorder is further enhanced.  Since we have OBC in the $y$ direction, such edge channels are formed on the lower and upper boundaries as well (see bottom panels). However, at very strong disorder, Anderson localization  appears, and surface states with an on-site potential much larger than the hopping amplitude effectively decouple from the bulk with the result that an effective disorder in the first inward layer is induced~\cite{SFFV12}.      
\subsection{TI with magnetic disorder} 
\begin{figure}[t]
\centering
\includegraphics[width=0.75\linewidth]{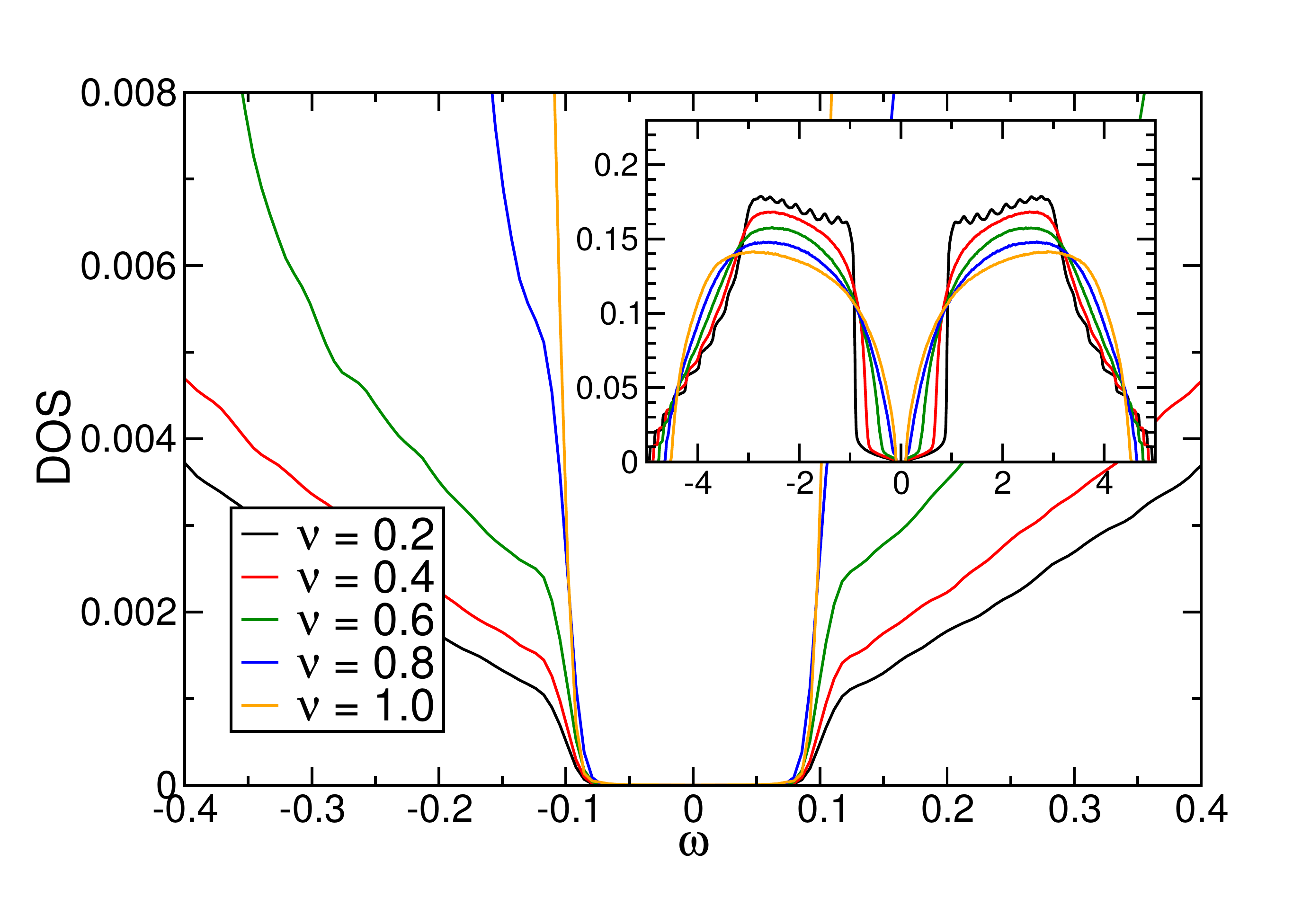} 
\includegraphics[width=0.89\linewidth]{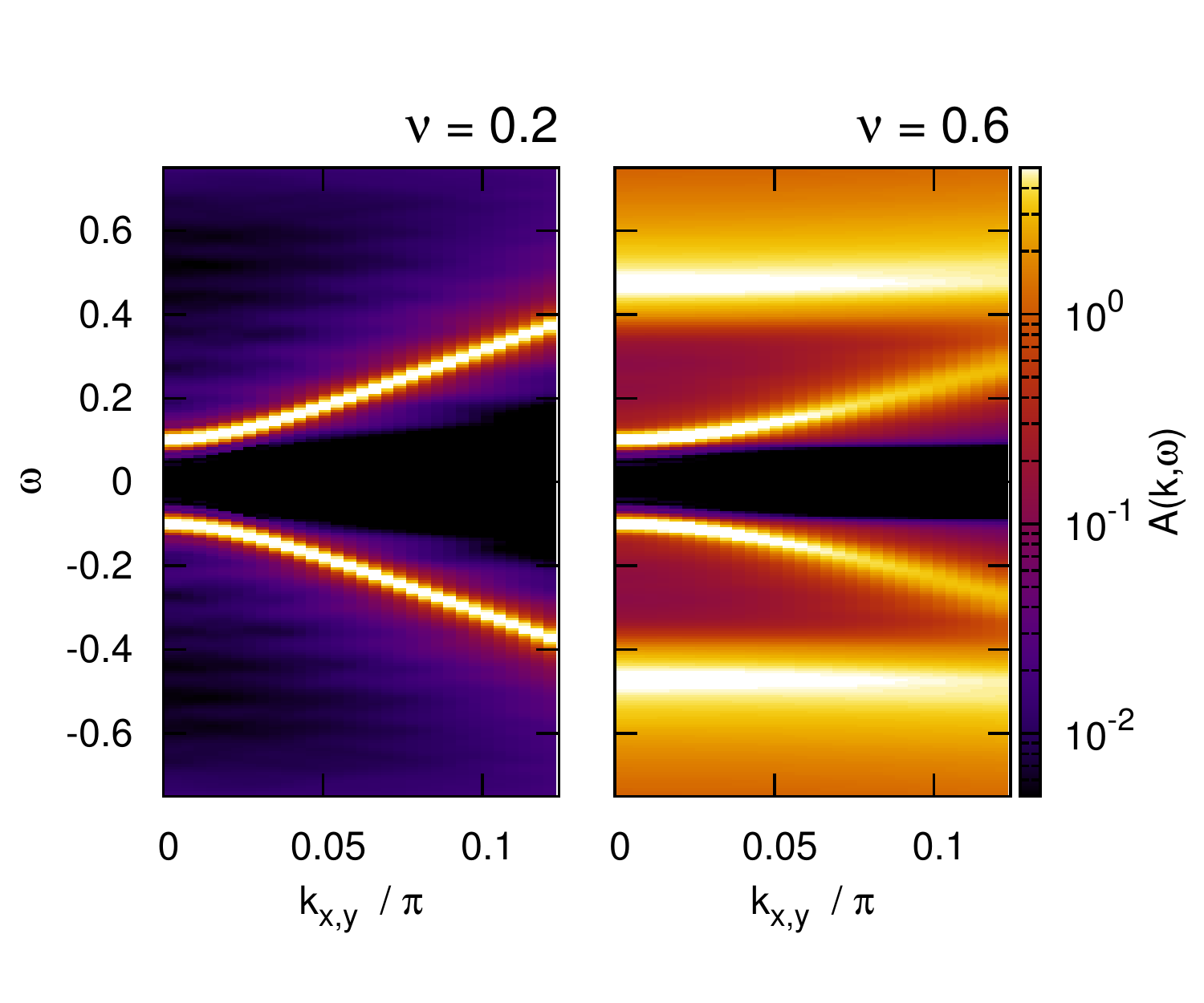} 
\caption{
(Color online)
 DOS (top) and spectral function (bottom) for a strong TI ($m=2$) with orbital disorder described 
  by a fluctuating Peierls factor  $\propto \ee^{\ii \tau_{n,j}}$. We consider a $(512\times 512 \times 10)$--site system with PBC and OBC
  and $\Delta_1=0.1$.  Again, the $H^Z$ term~\eqref{eq:TB_H_BZ} is neglected.
}
\label{fig8}
 \end{figure}

In this section we first address the orbital effects of a random magnetic field modeled by  a fluctuating Peierls phase factor attached 
to the  electronic transfer amplitude.  Figure~\ref{fig8} shows the DOS and single-particle spectral function at various 
disorder strengths. In comparison with Fig.~\ref{fig5}  the gap induced by $\Delta_1$ is less affected by  
bulk orbital disorder (simply because we have no shift of the local potentials, 
only a modulation of the hopping, i.e., of the electronic bandwidth).  Accordingly, the $\Delta_1$-gap    
shrinks somewhat in magnitude but persists. Apart from that, enhancing the disorder,  a lot of spectral weight (DOS) is transferred
from the  bulk states into the bulk pseudo-gap. There, as a new feature, a strong, almost dispersionless signal evolves.   

\begin{figure}[t]
\centering
\includegraphics[width=0.75\linewidth]{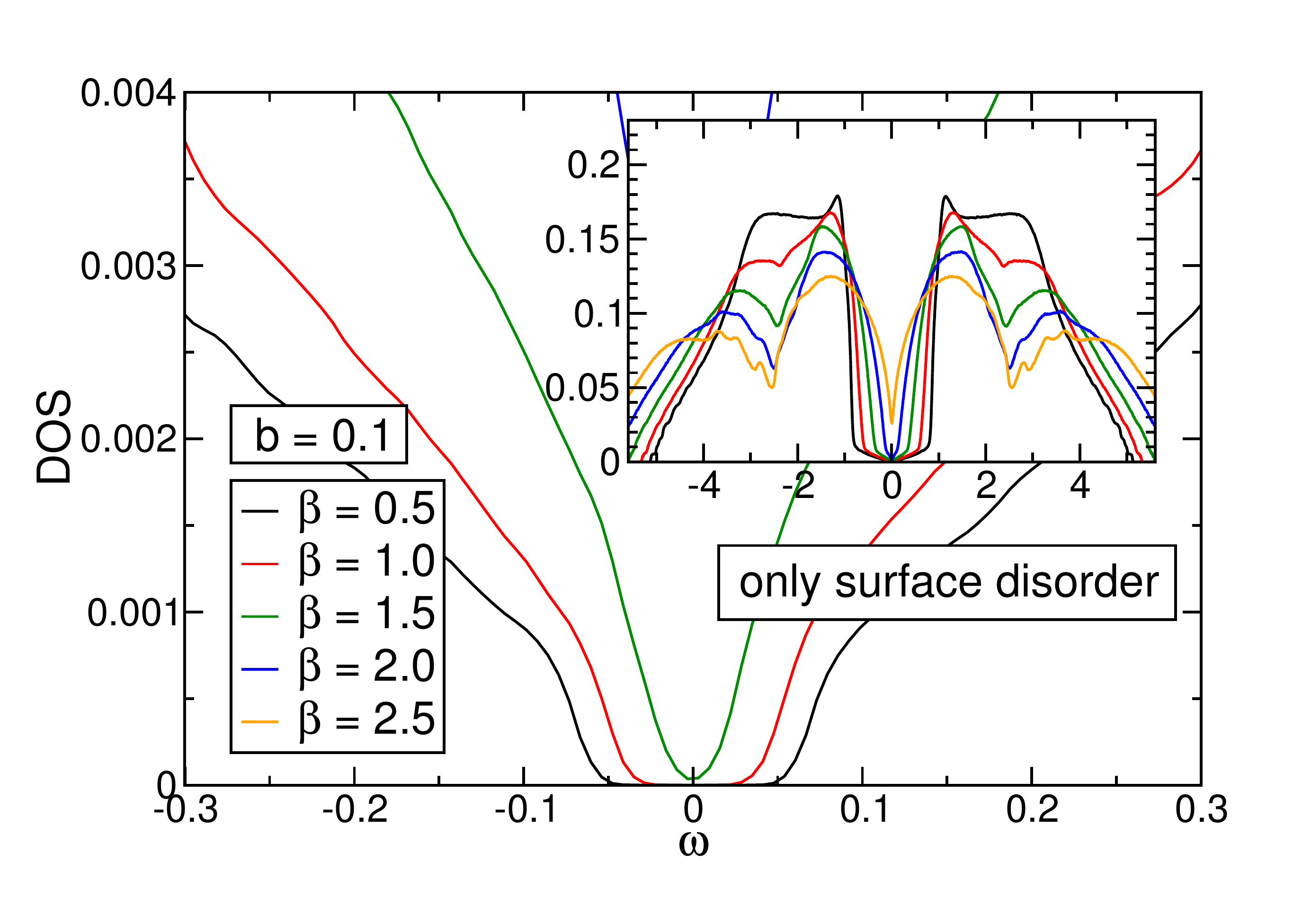} 
\includegraphics[width=0.89\linewidth]{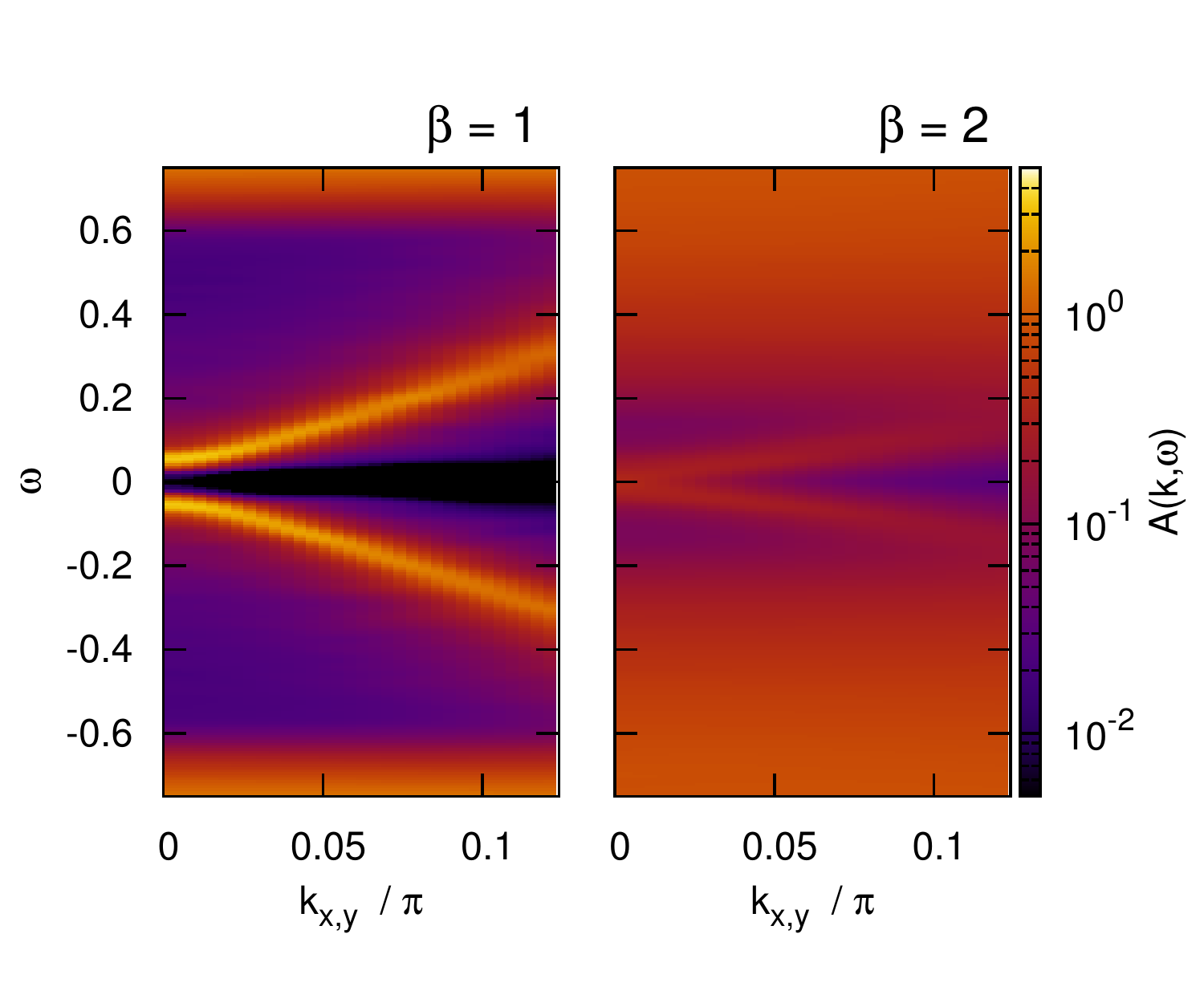} 
\caption{
(Color online) 
   DOS (top) and spectral function  (bottom) for a strong TI ($m=2$) with ``magnetic impurities'' only on the surface, 
   modeled by a random Zeeman term~\eqref{eq:TB_H_BZ} with $g^+ = 1$, $g^- = 0$. We zeroized $\Delta_1$, $\Delta_2$, $V_n$, and the $\tau_{n,j}$. 
}
\label{fig9}
\end{figure}

Next, we demonstrate that magnetic impurities on the surface of a TI will rapidly destroy the topological protected bands.
To this end, we set $V_n=0$, $\tau_{n}=0$, and  $\Delta_1=\Delta_2=0$  in our model Hamiltonian, and 
consider  only a random Zeeman term $H^Z$, where $g^+=1$ and $g^-=0$ without loss of generality. 
Here the gap in the surface state band structure is induced by a finite mean value $b = \langle B^Z_z \rangle$. Figure~\ref{fig9} illustrates the radical spectral weight (DOS) transfer into the bulk gap when  magnetic surface disorder is present. In addition two notches in the DOS appear at $\omega\simeq \pm 2.5$ (the DOS even vanishes at these points for magnetic bulk disorder with $1.9<\beta<2.8$; for still larger $\beta$ Anderson disorder sets in). Note that already for a rather moderate disorder strength ($\beta=2$) the midband gap closes and the photoemission spectrum becomes completely incoherent.  
\begin{figure}[t]
\centering 
\includegraphics[width=0.7\linewidth]{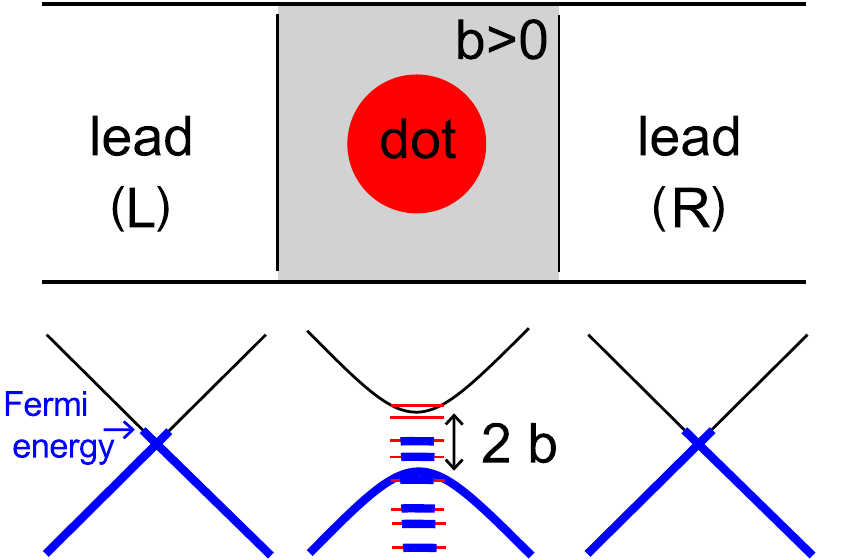} 
\caption{
(Color online) Schematic representation of a lead-contacted TI with a circular gate-defined quantum dot (red), which induces quasi-localized states in the surface-state band gap produced by a finite $b$. In the gated region $V_{dot}>0$. 
}
\label{fig10}
\end{figure}

\subsection{TI with a gate-defined quantum dot}
Finally, we analyze the capabilities of a controlled modification of transport through gate-defined nanostructures imprinted on the TI's surface. This can be achieved, just like  for graphene nanoribbons~\cite{FHP15}, by applying nanoscale top gates. That way a circular quantum dot can be created
that causes quasi-bound states inside the gap of the TI surface states produced by $b>0$ (see Fig.~\ref{fig10}).  In the following we consider a contacted strong TI sample with $32\times 32\times 6$ sites and PBC in the $y$ direction; the quantum dot has $R_{dot}=8$ (in units of the lattice constant). Within the leads we have $b=0$.
   
\begin{figure}[t]
\centering 
\includegraphics[width=0.8\linewidth]{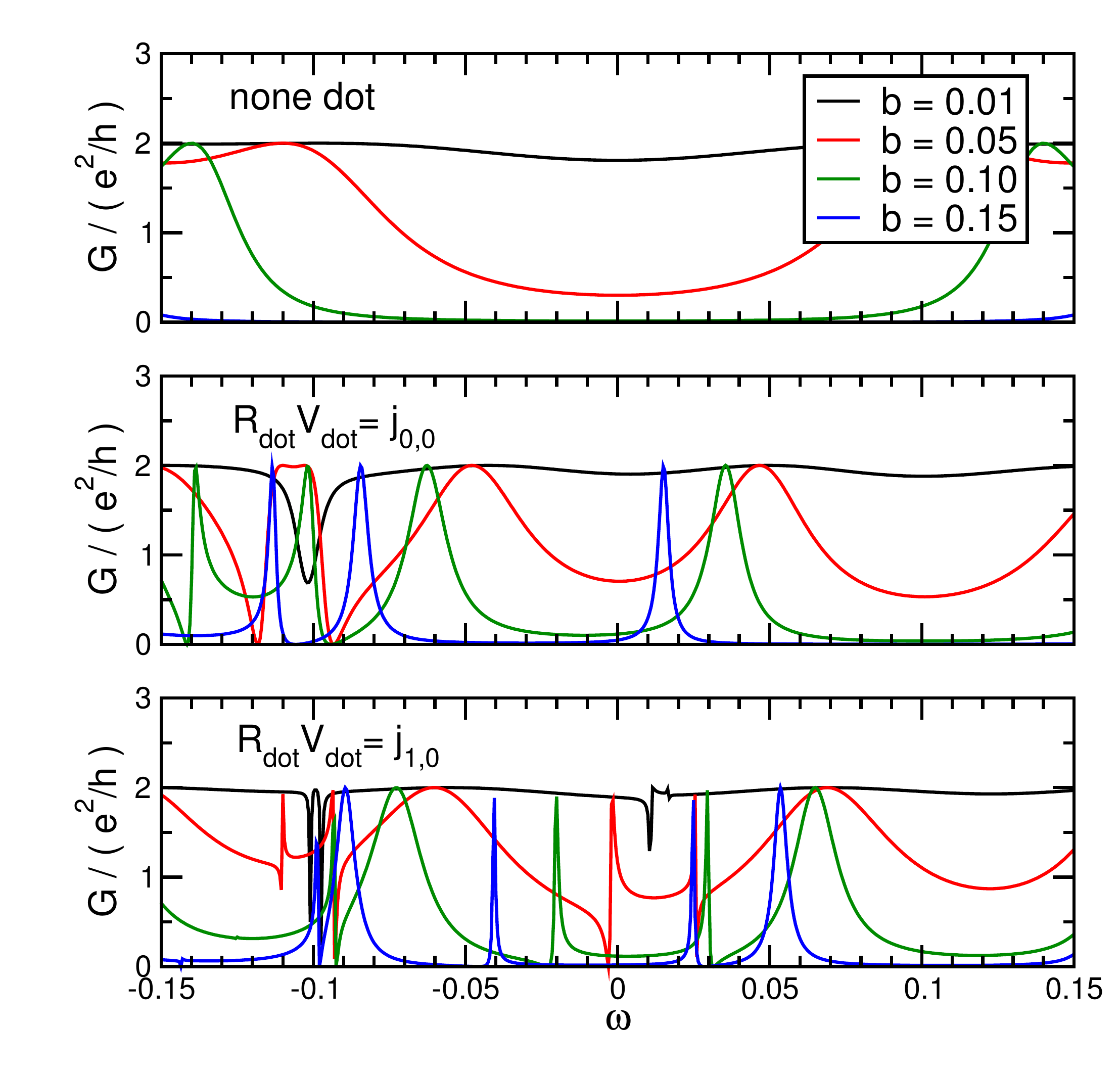} 
\caption{(Color online) Conductance of a contacted strong TI ($m=2$) in a homogenous magnetic field $b=\langle B^Z_z\rangle$.  
 For comparison only, the top panel  gives $G$ without imprinting a quantum dot. In the middle and lower panels the dot potential $V_{dot}$ is chosen such that the product $R_{dot} V_{dot}$ equals  $j_{0,0}\simeq 2.4$ and $j_{1,0}\simeq 3.8$, i.e., $V_{dot}=0.3006$ and  $V_{dot} = 0.47896$, respectively.}
\label{fig11}
\end{figure}

Figure~\ref{fig11} gives the conductance for such a setup.  In the case $V_{dot}=0$ (no dot) $G$ is reduced in the vicinity of the band center because of the gap triggered by $b$ (see top panel).  The maximum value for $G$ tells us that our finite system develops two open transport channels at most.  When a quantum dot exists, a series of resonances appears in the gap region owing to possible excitation of dot normal modes. Within the Dirac (continuum)  approximation for massless fermions these scattering resonances occur for particular combinations of $R_{dot}$ and  $V_{dot}$~\cite{HBF13a}. For an equilibrium situation (without incident wave) the normal modes can be interpreted as decaying states, where, for small values of $\omega$, the lifetime of these quasi-bound dot states (appearing for $R_{dot} V_{dot}=j_{n,m}$, where $j_{n,m}$ denotes the $m$-th zero of the $n$-th
Bessel function of the first kind)  can be extraordinarily long. This has been confirmed for a discrete (graphene) lattice by exact diagonalization~\cite{PHF13,PHWF14}. As can be seen from the middle and bottom panels of Fig.~\ref{fig11}, in our case the resonances can actuate  resonant tunneling. In a sense they act as doorway states~\cite{FWLB08}. Interestingly, therefore even the maximum possible value of $G$ can be achieved. When the gap is small the dot-bound states will hybridize with extended states. Accordingly, much broader peaks in  $G$ emerge than for a large-gap situation (large values of $b$).  Higher dot-bound modes narrow the $G$-signal as well. Note that the conductance undergoes a dramatic change if $\omega$ (or, alternatively,  $V_{dot}$) is slightly varied near the resonance points. In this way, such a system may act as a switch.
\begin{figure}[t]
\centering 
\includegraphics[width=0.9\linewidth]{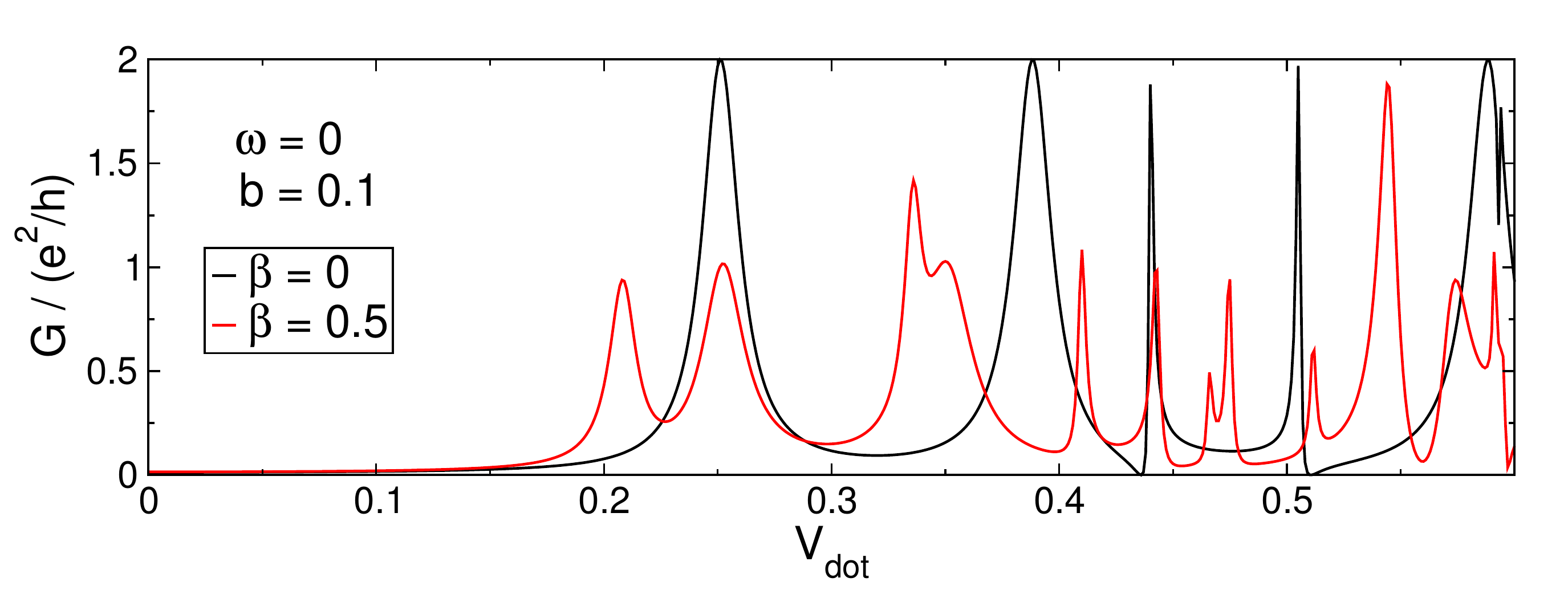} 
\caption{(Color online) Conductance of a strong TI ($m=2$) with a quantum dot ($R_{dot} = 8$) imprinted on the top surface. 
The system is subjected to a (random) magnetic field $B^Z$ with offset $b=0.1$ and $\beta=0$ ($\beta=0.5$); see black (red) curves.
For the disordered case ($\beta=0.5$), data for $G$ were averaged over ten samples.
}
\label{fig12}
\end{figure}

\begin{figure}[t]
\centering 
\includegraphics[width=0.53\linewidth]{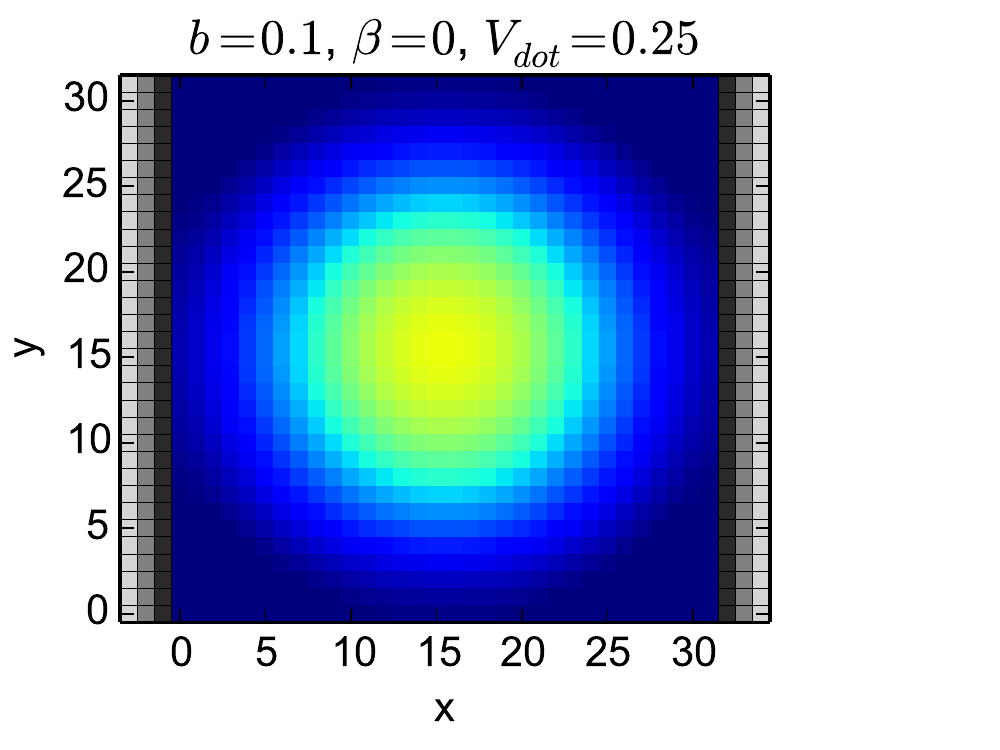}
\hspace*{-1.2cm} 
\includegraphics[width=0.53\linewidth]{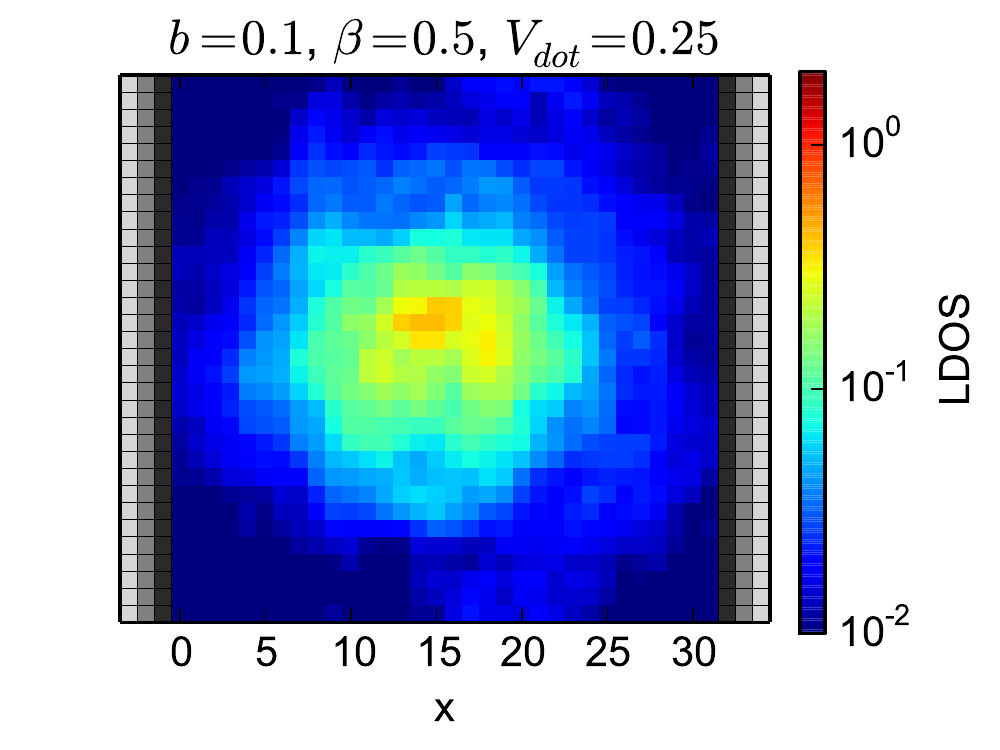}\\
\includegraphics[width=0.53\linewidth]{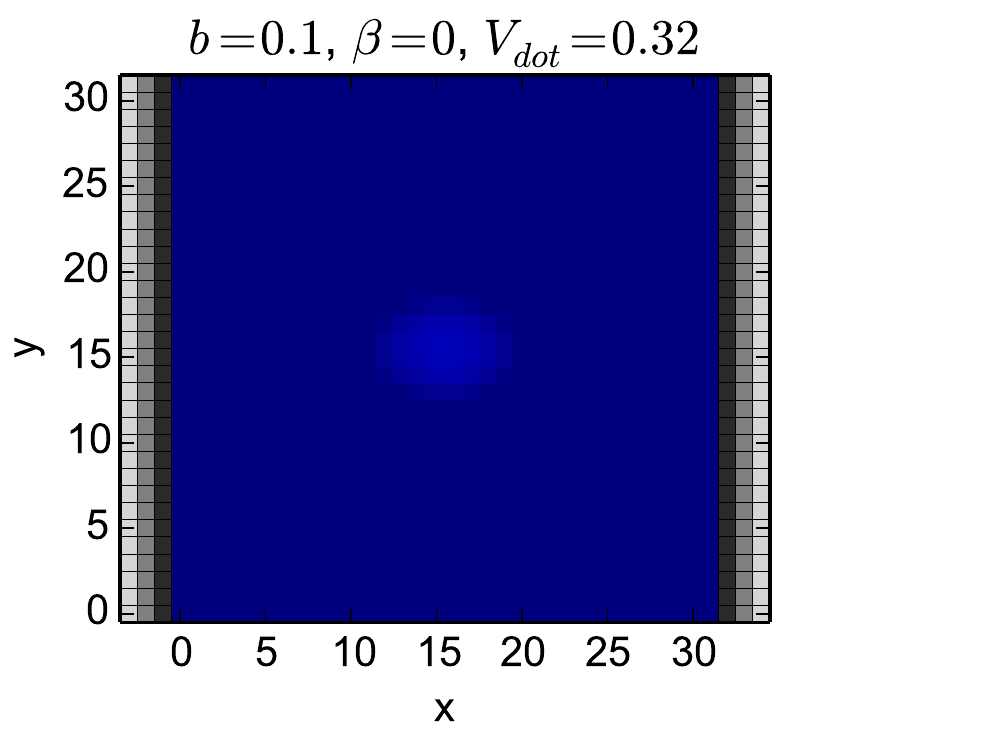}
\hspace*{-1.2cm} 
\includegraphics[width=0.53\linewidth]{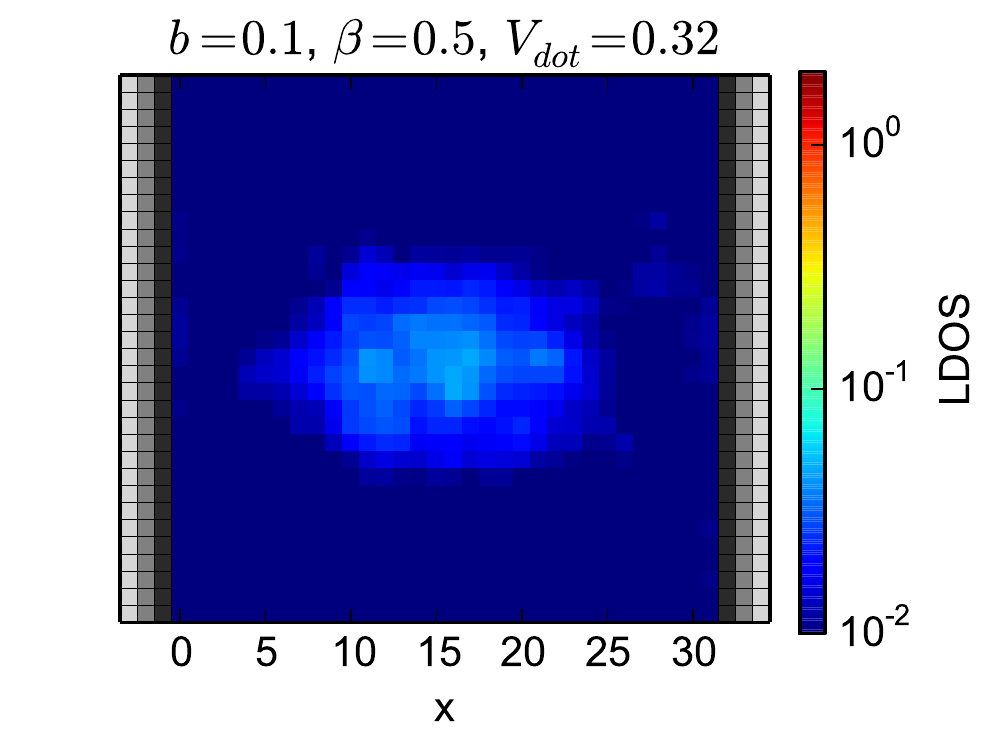} \\
\includegraphics[width=0.53\linewidth]{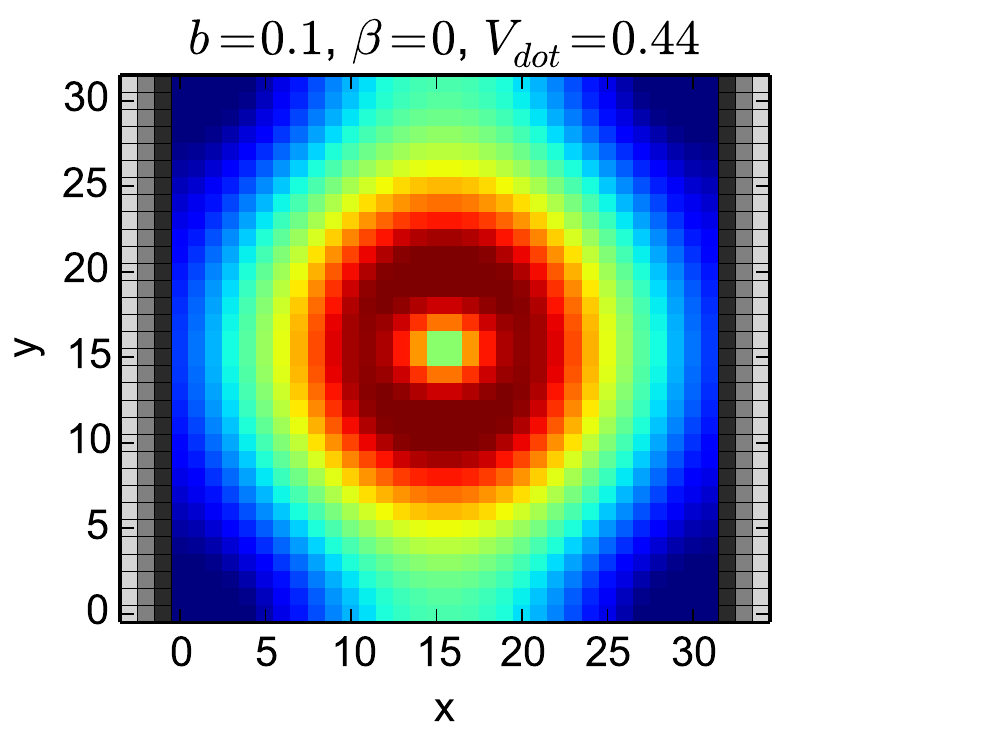}
\hspace*{-1.2cm} 
\includegraphics[width=0.53\linewidth]{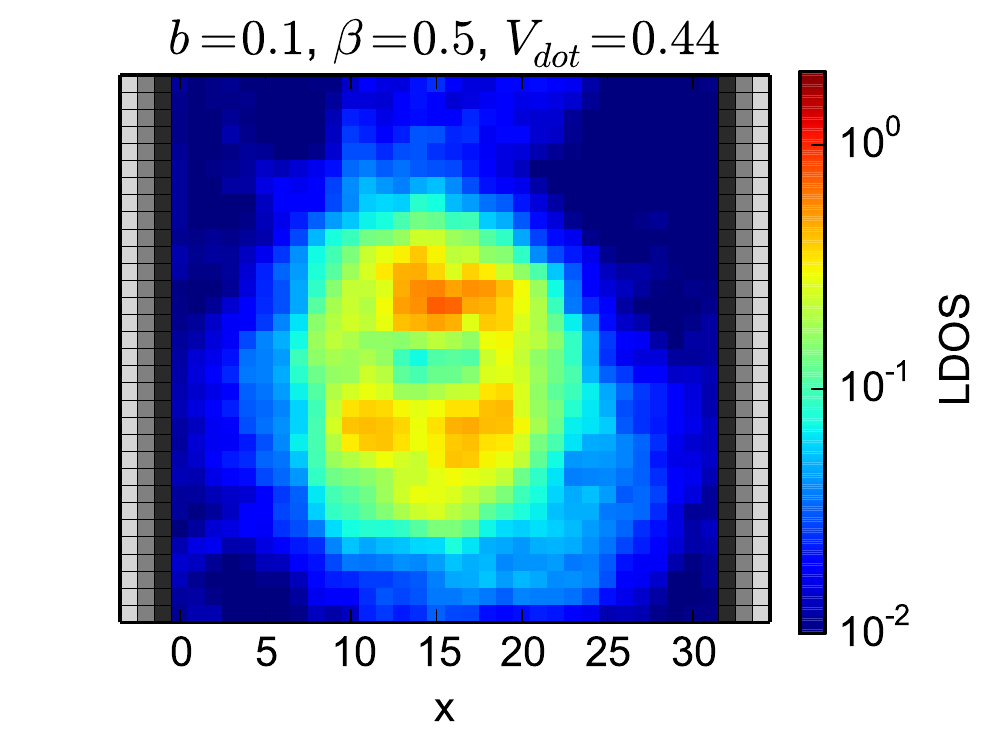} 
\caption{
(Color online)
LDOS at $\omega=0$ for a TI with a gate-defined  quantum dot  ($R_{dot} = 8$) on the surface.
Here $V_{dot} = 0.25$, 0.44 (0.32) were chosen in order to realize a resonance (a minimum) in the conductance 
(see Fig.~\ref{fig12}). The left (right) panels give the LDOS without (with) random magnetic fields.}
\label{fig13}
\end{figure}

\begin{figure}[h]
\centering 
\includegraphics[width=0.9\linewidth]{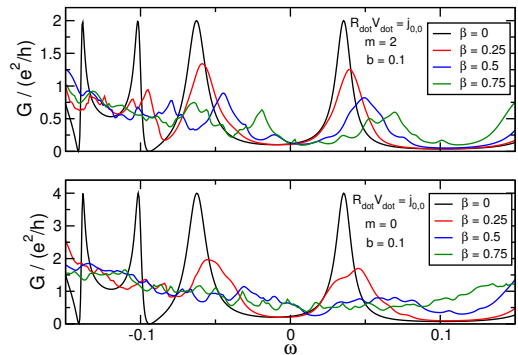} 
\caption{(Color online) Conductance of a strong (top) and weak (bottom) TI  with a gate-defined quantum dot ($R_{dot} = 8$,  
$V_{dot} = 0.3006$)  subjected to a random magnetic field $B^Z$ with offset  $b=0.1$. $G$-data are averages over ten samples.
}
\label{fig14}
\end{figure}

Magnetic disorder will split, shift, and partly suppress the conductance maxima (see Fig.~\ref{fig12}). The general features
of the conductance, including the resonant tunneling via quasi-bound dot states, persist, however. That means the tunability 
of transport is guaranteed even for (moderate) magnetic disorder. Figure~\ref{fig13} shows the LDOS for three representative gate-potentials without (left-hand panels) and with (right-hand panels) random magnetic potentials. In the top panels $V_{dot}=0.25$ is chosen to realize the dot mode lowest in energy. The spatial localization of the $\omega=0$ surface states is obvious. Remarkably, the corresponding conductance equals those for a junction without a quantum dot (see Fig.~\ref{fig12}). Disorder softens the circular shape, but the resonance still exists. That is why the conductance is finite, albeit reduced. At $V_{dot}=0.32$ the quantum dot states are out of resonance. Consequently, the LDOS on the surface almost vanishes. Here disorder may induce some states with the result that the conductance slightly increases. A higher dot bound mode (with larger orbital momentum) is implemented by $V_{dot}=0.44$. It possesses a ring-shaped  LDOS intensity, which is again weakened by disorder. Figure~\ref{fig14}  gives the conductance  depending on $\omega$, for strong and weak TI with a $b$-field induced gap, where $R_{dot} V_{dot}=j_{0,0}$.
Having two Dirac cones instead of a single one, the conductance of the weak TI is twice as large as for the strong TI. 
Since $V_{dot}>0$, $G$ is not symmetric with respect to $\omega \to -\omega$. 
On account of disorder ($\beta$) the resonances will be suppressed and $G$ is smeared out. This effect is more pronounced for weak TI because of the scattering between the two Dirac cones.

\section{Conclusions}
To summarize, the electronic properties of strong (and also weak) topological insulators are dramatically affected by external magnetic fields that break the inversion symmetry and time-inversion symmetry. The resultant gap formation causes massive Dirac fermion surface states. Both nonmagnetic and magnetic impurities (modeled by diagonal random potentials and Zeeman fields, respectively), but not orbital non-diagonal disorder, induces states into  this midband gap, yielding diffusive metallicity at the surface. Even so, the calculated angle-resolved photoemission spectra indicate that the surface states largely retain their bare dispersion, up to the point where disorder with a strength comparable to  or larger than the bulk gap leads  to Anderson localization.  From an application-technological point of view, the tunability of the transport properties of TIs by external electric and magnetic fields is of particular importance. We showed that quantum dots can be engineered on the TI's surface by nanoscale top-gates and can be used to control the conductance, meeting device requirements. 
\acknowledgments
The authors would like to thank  L. Fritz, R.~L. Heinisch, G. Schubert,  J. Tworzydlo, M. Vojta, and G. Wellein for valuable discussions.
A.P.  was funded by the Deutsche Forschungsgemeinschaft through the Priority Programme 1459 `Graphene' 
and by the Competence Network for Scientific High-Performance Computing in Bavaria (KONWIHR III, project PVSC-TM).

\bibliography{ref} 
\bibliographystyle{apsrev}

\end{document}